\documentclass[aps,preprint,showpacs ,amssymb, amsmath,pre]{revtex4}
\usepackage{bm}
\usepackage{graphicx}

\begin{document}

\newcommand{\mi}[1]{\mathindent #1 mm}
\newcommand{\vpr}[1]{\vec {p^{\prime}_{#1}}}
\newcommand{\bdm}{\begin{displaymath}}
\newcommand{\edm}{\end{displaymath}}
\newcommand{\beq}{\begin{equation}}
\newcommand{\eeq}{\end{equation}}
\newcommand{\beqa}{\begin{eqnarray}}
\newcommand{\eeqa}{\end{eqnarray}}
\newcommand{\hi}{\^{\i}}
\newcommand{\Sch}{Schr\"odinger }
\newcommand{\Schv}{Schr\"odinger}
\newcommand{\tc}{\textcolor}
\newcommand{\edo}{\end{document}}
\newcommand{\npb}{\nopagebreak}
\newcommand{\ssk}{\smallskip}
\newcommand{\msk}{\medskip}
\newcommand{\bsk}{\bigskip}
\newcommand{\mbx}{\mbox{\tiny x}}
\newcommand{\ds}{\displaystyle}
\newcommand{\vs}{\vspace{0.1cm}}
\newcommand{\np}{\newpage}
\newcommand{\bit}{\begin{itemize}}
\newcommand{\eit}{\end{itemize}}
\newcommand{\ben}{\begin{enumerate}}
\newcommand{\een}{\end{enumerate}}
\newcommand{\vem}{\vspace{1em}}
\newcommand{\hem}{\hspace{1em}}
\newcommand{\noi}{\noindent}
\newcommand{\rar}{\rightarrow}
\newcommand{\bb}{{a}}
\newcommand{\bbs}{{a }}
\newcommand{\Bb}{{A}}
\newcommand{\Bbs}{{A }}
\newcommand{\bc}{{q}}
\newcommand{\bbb}{{\~a}}
\newcommand{\bbbs}{{\~a }}
\newcommand{\bx}{{z}}
\newcommand{\cb}{{1}}
\newcommand{\bfP}{{\bf P}}
\newcommand{\bfp}{{\bf p}}
\newcommand{\bfA}{{\bf A}}
\newcommand{\bfB}{{\bf B}}
\newcommand{\bfC}{{\bf C}}
\newcommand{\balpha}{{\mbox{\boldmath$\alpha$}}}
\newcommand{\bfga}{{\mbox{\boldmath$\gamma$}}}
\newcommand{\bfE}{{\bf E}}
\newcommand{\bfe}{{\bf e}}
\newcommand{\bfF}{{\bf F}}
\newcommand{\bfL}{{\bf L}}
\newcommand{\bfR}{{\bf R}}
\newcommand{\bfS}{{\bf S}}
\newcommand{\bfa}{{\bf a}}
\newcommand{\bfd}{{\bf d}}
\newcommand{\bfn}{{\bf n}}
\newcommand{\bfkp}{{\mbox{\boldmath$\kappa$}}}
\newcommand{\bfsg}{{\mbox{\boldmath$\sigma$}}}
\newcommand{\bfSg}{{\mbox{\boldmath$\Sigma$}}}
\newcommand{\bfal}{{\mbox{\boldmath$\alpha$}}}
\newcommand{\bfro}{{\mbox{\boldmath$\rho$}}}
\newcommand{\bfr}{{\bf r}}
\newcommand{\bfAr}{{\bf A}(\bfr)}
\newcommand{\hbr}{\hat{\bf r}}
\newcommand{\bfJ}{{\bf J}}
\newcommand{\bfs}{{\bf s}}
\newcommand{\bfk}{{\bf k}}
\newcommand{\hbp}{{\hat{\bf p}}}
\newcommand{\hbk}{\hat{\bf k}}
\newcommand{\bfq}{{\bf q}}
\newcommand{\bfV}{{\bf V}}
\newcommand{\bfv}{{\bf v}}
\newcommand{\bnabla}{{\bf \nabla}}
\newcommand{\slm}{\sum_{l=0}^{\infty}\sum_{m=-l}^{l}}
\newcommand{\slmp}{\sum_{l'=0}^{\infty}\sum_{m'=-l'}^{l'}}
\newcommand{\viq}{\, , \qquad}
\newcommand{\ra}{\,\rangle}
\newcommand{\la}{\langle \,}
\newcommand{\tg}{\mathrm{\,tg\,}}
\newcommand{\cotg}{\mathrm{\,cotg\,}}
\newcommand{\arctg}{\mathrm{\,arctg\,}}
\newcommand{\ai}{{\^{\i}}}
\newcommand{\cU}{{\cal U}}

\title{Electron distributions in nonlinear Compton scattering}

\author{Madalina Boca}

\email{madalina.boca@g.unibuc.ro}
\affiliation{Department of Physics  and Centre for Advanced Quantum Physics,  University of Bucharest, MG-11, Bucharest-M\u agurele, 077125  Romania}

\author{Victor Dinu}
\affiliation{Department of Physics  and Centre for Advanced Quantum Physics,  University of Bucharest, MG-11, Bucharest-M\u agurele, 077125  Romania}

\author{Viorica Florescu}
\affiliation{Department of Physics  and Centre for Advanced Quantum Physics,  University of Bucharest, MG-11, Bucharest-M\u agurele, 077125  Romania}

\begin{abstract} 
 Based on quantum theory, we investigate the distribution of the electrons scattered in nonlinear Compton   effect by an electromagnetic plane wave.   The monochromatic case, examined in detail, reveals   features of the electron distribution,  useful in the understanding of the pulsed plane wave case for
 particular intensity and electron energy regimes.   The graphs displayed  focus on the case of  head-on or near head-on collision of an energetic electron  with an electromagnetic circularly polarized pulsed  plane wave and show that the deviation in direction is  extremely small, while   the  distribution in energy can  be  visibly different from that of the initial electron.  Two pulse shapes,  several laser intensities and  high incident electron energies are considered. 
\end{abstract} 

\maketitle

\section{Introduction}\label{s-i}

 Nonlinear Compton scattering (NLCS) is one of the simplest 
processes  predicted by quantum theory \cite{NNR,
SaHa,KrKa} and was  detected  in the  head-on collision  of an energetic electron beam  with an intense laser beam \cite{E144-1,E144-2}.     

In the case of a pulsed electromagnetic plane wave (a plane wave with a finite extension in the direction of propagation), the model we use for the laser beam,   an electron distribution at the end of the pulse different from the initial   one can  not be  predicted within classical electrodynamics (CED)  without including the radiation reaction (RR) as, according to this theory,   the pulse leaves each electron with the same momentum it had at the beginning of the pulse. On the contrary, the   emission of radiation   can be described by CED  as Thomson scattering: the  charged particle accelerated by the electromagnetic field  emits radiation during its  well determined motion.   The inclusion of  RR in CED takes into account the energy lost by the electron and leads to a final  electron momentum different of the initial one.  The classical description of RR  was analyzed  in several publications  in which not only the formalism was  discussed but also the effects of radiation reaction on Thomson scattering  (references can be found in the   very recent review \cite{RMP}). In quantum theory RR was considered only  very recently \cite{RR},\cite{Sok}.  It is argued that  the standard description of NLCS, as a  single   one-photon emission, using the Volkov solutions of Dirac equation, does not include radiation reaction and that the mechanism for it is the incoherent multiple  one-photon emission by the electron.

 Up to now   comparisons between quantum and classical predictions were done only for the emitted radiation spectrum.  In the work presented in this paper we do no not include RR effects.

 In {\it quantum theory} NLCS is described as the {\it  spontaneous emission of  one photon} by the electron interacting  with an intense external electromagnetic field.  The electromagnetic field is described classically,  an approximation which is justified for high intensity; the most used model until recently  was  the monochromatic plane wave. The theoretical studies published  in the last three years adopt a more realistic description of short pulses,  the pulsed plane wave model,   for NLCS \cite{BF,altii1,altii2,altii3,altii4} as well for  other processes \cite{AI,altele}.  The finite transverse extension of a real laser beam  is neglected in quantum calculations, where an adequate treatment of this aspect  was not developed up to now. On the contrary,   in   calculations based on CED, where it is possible to include any shape of the electromagnetic field,   beam size effects have been  already considered \cite{size,galkin}.

  From a  predicted  probability distribution for the simultaneous detection of the emitted photon and  the scattered electron, quantum theory extracts separate information on  the emitted radiation, to which the majority of the results in the literature refer, or on  the electron distribution, less studied up to now.  It was remarked \cite{altii1} that  in experiments the radiation emitted in NLCS was not investigated, contrary to the Thomson case where the angular distribution of  first several  harmonics has been   recorded  \cite{Th-exp}. 
 Electrons have been detected  in the E-144    experiment at SLAC performed 15 years ago: the collision   of a 46.6 GeV electron beam with 
terawatt pulses from a Nd:glass laser at 1054 and 527 nm wavelengths. Comparison with the theoretical energy spectra presented in Fig. 4 of \cite{E144-1} reveals the absorption of up to four laser photons. More than that, a suplementary evidence was  given by the detection of positrons \cite{E144-2} which come from   a succession of two elementary processes: NLCS,  and $(e^-,e^+)$  pair creation by the energetic photon emitted in the first process.

    In the monochromatic case, the electron energy distribution  presents thresholds \cite{E144-1,E144-2} that can be connected with the absorption of fixed number of laser photons; these thresholds will be discussed in Sect. \ref{s-iii}. 

In connection with the perspective of very intense sources of radiation \cite{int}, the interest of studying  NLCS along with other processes in the very intense regime (intensities above 10$^{22}$ W/cm$^2$) has increased and it is reflected in the most recent works \cite{arki}.  Theoretical aspects as the use of a wave packet for the description of the initial electron  \cite{CP1} or the quantum description of the external electromagnetic field  are reanalyzed \cite{CP2}. 

 Our paper is a theoretical study of the {\it electron distribution in NLCS} occurring  in the  interaction of the electron with a laser pulse.  We have recently published  a few results from a preliminary investigation of this distribution \cite{BDF}.  The present paper  is organized as follows. Section \ref{s-ii} displays the general expression of the multiple differential  distribution describing  both photons and electrons, from which analytic expressions for the electron distribution in the monochromatic or pulse case result. The monochromatic case is  discussed in Sect. \ref{s-iii}.  Our purpose is not a numerical calculation  of the transition rates, but an analysis of the position of the singularities they  present in this ideal case. In the study of  the  electron distributions in the monochromatic case, one has to distinguish between the  "bare momentum" $\,\bfp_2\,$  (the momentum of the asymptotically free electron) and the "dressed momentum" $\,\bfq_2\,$ of the scattered electron; these momenta are in biunivocal correspondence.  We have found that the analytic part of this exploration is more easily done in terms of the dressed momenta.  The electron distribution present   $\,\delta$-type singularities at particular  values of  the final dressed momentum $\,\bfq_2^{(N)}\,$, indexed by the positive integer $N$  which is interpreted as the number of laser photons absorbed by the electron.  We follow these singularities in the space of  the polar coordinates  ($\,\mid {\bf q}_2|,\theta_{q_2},\phi_{q_2}$)    of the vector $\,\bfq_2\,$. The manifold of the singularities   for a fixed $\,N\,$ is a surface  which may be closed or not. We find  a subset of points on this surface at which the distribution  has a particular type of singularity that influences  the electron angular distribution.  Then,  we  translate the results in terms of  the bare momentum ${\bf p}_2$  of the electron   and use them in the   numerical illustration  that concludes Sect. \ref{s-iii}.  We find that,  for not too high laser intensity,  this  type of  representation    is  useful  in understanding   the distributions obtained with  finite pulses.  Section \ref{s-iv} is devoted to  the equations valid in the pulse case.  The numerical results  presented in Sect. \ref{s-v}  for two type of pulses reveal conditions in which the analysis done in the monochromatic  case  is useful.

In  Appendices  \ref{s-a} and \ref{s-b}  we give  details about the  justification of some of the results   presented in Sect. \ref{s-iii}  concerning the monochromatic case.

\section{The theoretical framework. General expression for the transition probability} \label{s-ii}

The general theoretical framework is the same as in \cite{BF}, so not too many details will be given here. The same formalism is  described  in other recent publications \cite{altii1,altii2,altii3}.

The system investigated  consists in an electron (charge $\,e<0$, mass $\,m\,$) interacting with the quantized electromagnetic field  describing the emitted photon and with a classical electromagnetic plane wave with  the unit vector $\,{\bf n}_1\,$ in the  direction of propagation  and  described by a vector potential 
\beq
\bfA(\phi) \viq \phi=t-{{\bf n}_1\cdot\bfr/c=\frac{n_1\cdot x}{c}}\,,
\eeq
where $\,c\,$ is the velocity of the light  and $\,n_1\equiv(1,{\bf n}_1)$  is the notation for a four vector associated to the laser propagation direction. 
We  take the $z$ axis along $\,\bfn_1\equiv \bfe_z\,$ and we work with   the vector potential  
\beq\label{puls}
\bfA(\phi)=A_0\,f({\phi})\,[\,{\bf e}_x\,\cos(\zeta/2)\,\sin(\omega_1\,\phi)+{\bf e}_y\,\sin(\zeta/2)\,\cos(\omega_1\,\phi)\,]\,.
\eeq
This way the  unit  vectors $\,\bfe_x\,$ and $\,\bfe_y\,$ of the $x$ and $y$ axes  are chosen along the axes of the ellipse associated with the state of  polarization of the external field. The wave polarization is described by the parameter $\,\zeta\,$ ($\,\,\zeta=0\,$ and $\,\pi\,$ correspond to linear polarization, $\,\zeta=\pm \pi/2\,$ to circular polarization);  in  our numerical calculations only the case of circular polarization  will be considered.  In the {\it monochromatic case},  $\,f=1\,$ and   $\,\omega_1\,$ is the  laser frequency; in the case of a {\it laser pulse}, the function $\,f\,$ is the pulse envelope, supposed to be significantly different from zero only in a finite interval $\,(\phi_{\mathrm{in}},\phi_{\mathrm f}),$ and $\,\omega_1\,$ is the central frequency.   The maximum amplitude reached by  the electric field of the pulse is  $\,\omega_1\,A_0\,$  and the  electromagnetic field intensity is characterized by the  dimensionless parameter
\beq
\eta=\frac{\mid e\mid \,A_0}{mc}\,.
\eeq

The formalism we use starts with the general definition of the transition amplitude between two states of the  system  (electron + quantized  electromagnetic field + classical electromagnetic plane wave), 
\begin{equation}
{\cal M}_{1\rightarrow 2}=\lim_{t_2 \rar \infty}\, \lim_{t_1\rar-\infty}\langle \Psi_2(t_2)|U(t_2,t_1)|\Psi_1(t_1)\rangle\,,
\end{equation}
 with $\,U\,$  the evolution operator of the system. The initial and final states are products of free electron states of momenta ${\bf p}_1$ and ${\bf p}_2$ with, respectively, the vacuum state of the electromagnetic field and the one photon state of momentum $\,\bfk_2\,$ and polarization $\bfs_2$.  In contrast to \cite{BF}, were a spinor with well determined momentum,  normalized on an  arbitrary volume $\,V\,,$ was used for the initial state of the electron,  now we describe formally the electron with a  momentum $\,{\bf p}_1$ by a ``wave-packet'', 
\beq\label{pachet} 
\Psi_{1}(\bfr,t)=\int_{\bfp}\Phi(\bfp)\frac{e^{\frac{i}{\hbar}(\bfp\cdot\bfr-E\,t)}}{(2\pi\hbar)^{3/2}}\,\xi(p) \,d\bfp \viq \mid\Phi(\bfp)\mid^2=\delta(\bfp-\bfp_1)\,,
\eeq
 where $\,\xi(p)$ is a solution of Dirac equation $\,(\hat p-mc)\,\xi(p)=0
\,$ and it is normalized to 1 ($\,\xi^\dag\xi=1$), and $\,E=\sqrt{m^2c^4+c^2\bfp^2}$. This  procedure was  used recently in  \cite{CP1} for a spinless particle. 

The interaction  of the electron with the quantized electromagnetic field, responsible for the photon emission, is treated  in the  first order of  perturbation theory. The action of the free evolution operator $\,U_0\,$ (describing only the electron in the classical electromagnetic plane wave) on the free states leads to  the Volkov states  for which we use the explicit expression in  Eq. (B4) of \cite{BF}, with $\,V\,$ replaced by $\,(2\pi \hbar)^3\,$.  In  the  following formulas  we use the  four-momenta $\,p_1$, $p_2\,$ of the  initial, respectively final electron [\,$\,p_j\equiv(E_j/c,\bfp_j)\,,$  $\,E_j=c\sqrt{m^2c^2+\bfp_j^2},\;j=1,2$\,] and $\,k_2\,$ of the photon.

  The  expression of the  {\it transition probability}  for the emission of a photon with the wave-vector  $\,{\bf k}_2 \in d{\bf k}_2$ and a scattered electron with momentum $\,\bfp_2\in d\bfp_2\,$, averaged over the initial spin of the electron and summed over the final spin, is : 
\beq\label{s4}
d^4\Pi_{\mathrm{unpol}}
=\Pi_4(k_2,p_2)\,\delta({\bf p}_{1\perp}-{\bf p}_{2\perp}-\hbar {\bf k}_{2\perp})\,\delta[ n_1\cdot( p_1-p_2-\hbar k_2)]\,d{\bf k}_2\,d{\bf p}_2\,,
\eeq
where the subscript $\,\perp$ is used to indicate the components orthogonal on the laser propagation direction ${\bf n}_1$   and the function $\,\Pi_4\,$ has the expression
 \begin{eqnarray}\label{Pi4}
&&\Pi_4(k_2,p_2)=\frac{e_0^2}{4\,\pi^2}\,\frac{m^2c^5}{E_{2}\,(n_1\cdot p_{1})}\,\frac{1}{\hbar\omega_2}\left\{|{\cal B}|^2\left(-1+\frac{(p_1\cdot n_1)(p_2\cdot k_2)+(p_1\cdot k_2)(p_2\cdot n_1)}{(mc)^2k_2\cdot n_1}\right)+\right.\nonumber\\
\quad &&\left(1+\frac{(\hbar k_2\cdot n_1)^2}{2(n_1\cdot p_1)(n_1\cdot p_2)}\right)\left[|{\mbox{\boldmath${\cal{A}}$}}|^2-2\frac{(n_1\cdot p_1)(n_1\cdot p_2)}{(mc)\,\hbar k_2\cdot n_1}\left.\Re\left\{ {\cal B}^*\left(\frac{{\mbox{\boldmath${\cal{A}}$}}\cdot {\bf p}_1}{n_1\cdot p_1}-\frac{{\mbox{\boldmath${\cal{A}}$}}\cdot {\bf p}_2}{n_1\cdot p_2}\right)\right\}\right]\right\}\,,\qquad
\end{eqnarray}
( $e_0=e/\sqrt{4\pi\epsilon_0}$ ).
The external field dependence is contained in three one-dimensional integrals $\,{\cal B},\,{\cal A}_x\,$ and $\,{\cal A}_y\,$ defined as 
\begin{eqnarray}
{\cal B}(2,1)&\equiv& \int\limits_{-\infty}^{\infty}d\phi\,
\exp\left[-\frac{\mathrm i}{\hbar}G(p_1,p_2,k_2;\phi)\right]\,,\label{integr}\\
{\mbox{\boldmath${\cal{A}}$}}(2,1)&\equiv&-\int\limits_{-\infty}^{\infty}d\phi\,\frac{e{\bf A}(\phi)}{mc}\,\exp\left[-\frac{\mathrm i}{\hbar}G(p_1,p_2,k_2;\phi)\right]\,,\label{integrA}
\end{eqnarray}
where the function $\,G(p_1,p_2,k_2;\phi)$ is 
\begin{equation}
G(p_1,p_2,k_2;\phi)=\left[\,c\frac{\phi}{2}\,\widetilde n_1\cdot(p_1-p_2-\hbar k_2)+F(p_1;\phi)-F(p_2;\phi)\,\right] \,,
\end{equation} 
\beq\label{F}
F(  p;\phi)=\frac{c}{2\, n_1\cdot p}\int\limits_{\phi_0}^{\phi}d\chi\,[\,e^2{\bf A}^2(\chi)-2e{\bf A}(\chi)\cdot{\bf p}\,]\,,
\eeq
  with  $\,\widetilde n_1\equiv(1,-{\bf n}_1)$.  In the case of a pulse, where the vector potential is different from 0  for $\,\phi \in (\phi_{\mathrm{in}},\phi_{\mathrm f})\,,$ one has  $\phi_0=\phi_{\mathrm{in}}\,$. In the monochromatic case the indefinite integral  can be used,  as the change of the  value given to  $\phi_0$ leads only to the modification  of a phase factor in the Volkov solution.

Finally, we  remind here a classicality criterion presented several times in the literature (see, for example, \cite{altii1,altii4}):  the scattering of the radiation can be treated in the framework of CED, if the ratio  $
(\,s_{\mathrm{eff}}\,\hbar\omega_1\,\gamma_1\,)/(\,m\,c^2\,(1+\eta^2)\,)\,$,
with $\,\gamma_1\,$ the Lorentz factor of the initial electron and $\,s_{\mathrm{eff}}\,$ the maximum number of laser photons absorbed, is small compared to 1.  For $\,\eta\le1$, $\,s_{\mathrm{eff}}\,$ is  of the order of unity; for $\,\eta\gg1\,$, $\,s_{\mathrm{eff}}\,$ increases rapidly, proportional to $\,\eta^3$, and the ratio becomes 
\begin{equation}
y=\frac{\eta\hbar\omega_1\gamma_1}{mc^2}\label{classic1}\,.
\end{equation}
If $\,y\,$ becomes of the order of unity or larger, then the quantum behaviour sets in, and, as discussed before, one can expect  to obtain a final electron distribution different from the initial one.

\section{The monochromatic case}\label{s-iii}

In the monochromatic case, as known for long time \cite{NNR}, the integrals 
$\,{\cal A}_x, {\cal A}_y\,$ and $\,{\cal B}\,$  have analytic expressions as series of generalized Bessel functions. In these series  each term contains  an  one-dimensional $\,\delta$-function, as illustrated here by the integral $\,{\cal A}_x\,$,
\begin{equation}\label{Axserie}
{\cal A}_x(2,1)=\sum\limits_{N=-\infty}^{\infty}
A_x^{(N)}\delta[\widetilde n_1\cdot(q_1+N\hbar k_1-q_2-\hbar k_2)]\,.
\end{equation}

The four-momentum $\,k_1\,,$ 
\begin{equation}
k_1\equiv(k_1^0,{\bf k}_1)=\frac{\omega_1}c n_1\,,
\end{equation}
 is interpreted as the momentum of a photon associated to the electromagnetic  monochromatic plane wave.  In Eq. (\ref{Axserie})  appears the dressed four-momentum  $\,q$, a quantity met also in   the description of the electron motion in classical theory,  connected with the bare four-momentum $\,p\,$ by   
\beq\label{relpq}
q=p+\frac{m\,U_P}{n_1\cdot p}\,n_1\viq    U_P=\frac{e^2\,A_0^2}{4\,m}\viq\,n_1\cdot p=n_1\cdot q\,.
\eeq
  Between the two  4-momenta  $\,q\,$ and $\,p\,$ the correspondence is biunivocal, $\,p\,$ can be expressed as a function of $\,q\,$ as
\begin{equation}\label{relqp}
p=q-\frac{m\,U_P}{n_1\cdot q}\,n_1.
\eeq
For a given four momentum $\,p$, the first component  is $\,p_0=E/c\,$  and  the first component of  $\,q\,$ is   $\,q_0=W/c\,$, where 
\beq\label{EW}
E=c\sqrt{\bfp^2+m^2\,c^2} \viq W=c\sqrt{\bfq^2+m_*^2\,c^2} \viq m_*=m\,\sqrt{1+\frac{e^2\,
A_0^2} {2\,m^2\,c^2}}\,,  
\eeq
with  $\,m_*\,$ the dressed-mass, named also   the shifted mass. 

 The use  in the fully differential probability (\ref{s4})  of the integrals $\,{\mbox{\boldmath${\cal{A}}$}}(2,1)\,$   and $\,{\cal B}
(2,1)\,$, as series of $\,\delta$-functions similar to (\ref{Axserie}), gives an expression for $\,d^4\Pi_{\mathrm{unpol}}\,$ from which, after standard manipulations,   one extracts  {\it the transition rate}, denoted $\,d^4\Gamma$. It has   the structure  
\beq\label{multi}
d^4\Gamma(\bfp_2,\bfq_2)=\sum_{N=1}^\infty \gamma_4^{(N)}({\bf q}_2,{ \bf k}_2)\,\delta(q_1+N\hbar k_1-q_2-\hbar k_2)\,d{\bf q}_2\,d\bfk_2\,,
\eeq
 i.e., $\,d^4\Gamma\,$ is  a  series of four-dimensional $\,\delta$-functions with coefficients depending on the variables $\,{\bf q}_2\,$ and $\,{\bf k}_2$.  A term with fixed $\,N\,$ in the previous expression is  the contribution to the differential rate of the process in which $\,N\,$ laser photons have been absorbed.

In the monochromatic case it is customary \cite{qdist} to present these distributions as functions of ${\bf q}_2$, but it is also possible to present them taking as variable ${\bf p}_2$, using the relation
\beq
 d\bfq_2=\left(1+\frac{mc\,
U_P}{(p_2\cdot n_1)\,E_2}\right)\,d\bfp_2\,,
\eeq
with the ponderomotive potential $\,U_P\,$ defined in (\ref{relpq}).

In the following we shall suppose that in the "partial rates"  $\,\gamma_4^{(N)}\,$ the connection between the momenta imposed by the $\,\delta$-function was observed. In fact   the product of four one-dimensional $\,\delta$-functions leaves   arbitrary only two of the six components of the three-dimensional final momenta  $\,\bfq_2\,$ and  $\,\bfk_2\,$. Our purpose is not the evaluation of the partial rates, but the analysis of the implication of the  conservation rules for each term  with fixed $N$, in the case of the electron distribution.   As we shall see in Sect. \ref{s-v}, in appropriate conditions, connections are possible between the results of this analysis and the electron distributions in the pulsed wave case. 

\vspace{0.1cm}

In order to get the differential rates describing the
{\it electron energy and angular distributions} we have to integrate 
 over the emitted photon momentum. This is a direct operation performed by three of the four $\,\delta$-functions in each term and it imposes the following values to the emitted photon momentum
\beq\label{cons}
\hbar\widetilde \bfk_2 = \bfq_1-\bfq_2+ N\hbar{\bf k}_1\,.
\eeq
 The expression allowed for  the frequency is  $\;c\,\widetilde k_2^0\equiv\,c \mid \widetilde \bfk_2\mid$. After the integration on $\,\bfk_2\,$  only  an one-dimensional $\delta$-function is left in each term of the series which represents the double   differential rate $\,d^2\Gamma_e\,$ describing the scattered electron,  in terms of the dressed momentum ${\bf q}_2$,
\beq\label{distr} 
d^2\Gamma_e=\sum_{N=1}^\infty \Gamma^{(N)}(\bfq_2)\,\delta\left[\frac1{m_*c}(q_2^0+\hbar\, \widetilde k_2^{0}-q_1^0-N\,\hbar k_1^0)\right]\, d{\bf q}_2\,, 
\eeq
with $\,\Gamma^{(N)}({\bf q}_2)\equiv \,\gamma_{4}^{(N)}({\bf q}_2,\widetilde{\bf k}_2)/(m_*c\hbar^3)$.

In the following we emphasize some particularities of the electron distributions that come out from an  analysis of the argument of the $\,\delta$-function in (\ref{distr}).  

\subsection{Simultaneous detection of electron energy and direction}\label{s-iiia}

In the monochromatic case the  final electron distribution is   written as a  series of $\,\delta$-functions, as displayed by Eq. (\ref{distr}).    We shall study the position of the singularities in the  variables  $\,W_2\,$  [ or, equivalently, $|{\bf q}_2|$ related to $W_2$ by (\ref{EW}) ], and
 $\,\widehat \bfq_2\,$, the unit vector along  the 
direction of the final dressed momentum  $\,\bfq_2\,$  of polar angles $\theta_{q_2}$ and $\phi_{q_2}$. It is convenient to  introduce a new four-vector
\beq\label{nota} 
Q_N=q_1+N\hbar k_1\equiv(\frac{{\cal E}_N}c,{\bf Q}_N),\quad {\bf Q}_N\equiv \bfq_1+N\,\hbar {\bf k}_1 \viq {\cal E}_N\equiv W_1+N\,\hbar \omega_1\,
\eeq
 for which we have
\beq\label{lenQ}
Q_N^2= (m_*c)^2+2\,N\hbar k_1\cdot q_1\,.
\eeq 
We emphasize that  for given laser intensity and fixed $N$,  the four-vector  $\, Q_N\,$ is well determined only by the values of $\,q_1\,$ and $\,k_1\,$.    In the reference frame described at the beginning of Sect. \ref{s-ii}, we denote  by $\,\theta_N\,$ and $\,\phi_N$ the polar angles of the vector $\,{\bf Q}_N$,  and by $\,\alpha_N\in[0,\pi]\,$ the angle between  the momentum $\,\bfq_2\,$ and the vector $\,{\bf Q}_N\,$.
With (\ref{nota})   the argument of the $\,\delta$-function  in a term with {\it fixed N} in  (\ref{distr}) is  
\beq\label{FN}
F_N({\bf q}_2)\equiv  \frac1{m_*c}(q_2^0+\hbar\widetilde k_2^0-Q_N^0)=\frac1{m_*c^2}(W_2+c\mid {\bf Q}_N- \bfq_2\mid-{\cal E}_N\,).
\eeq
 The equation 
\beq\label{FN0} 
F_N(\bfq_2)=0\,.
\eeq 
determines the position of the singularities in the space of the  variables 
 $\,W_2$ (or $\,\mid \bfq_2\mid\,$) and $\,\hat \bfq_2\,$.  The condition (\ref{FN0}), for $\,N\ge1$ defines a family of surfaces in the space $(|{\bf q}_2|,\,\theta_{q_2},\,\phi_{q_2})$; the differential rate (\ref{distr}) has a $\delta$-type singularity along these surfaces and is zero otherwise.

 Before  going in more detail,  we  draw attention to  an approximate symmetry property of $\,F_N\,$, valid  in the case  $\eta\sim1$, when the number of terms which gives practically non-negligible contribution to the  electron distribution is limited to a value $\,N_{max}\,$ of the order of unity.  In this case it makes sense to analyze the condition (\ref{FN0}) only for $\,N\le N_{max}$. If, in addition,  $\,\gamma_1\gg1\,$ and the direction of  the bare momentum $\,{\bf p}_1\,$ is not too close to $\,{\bf n}_1$, then,  the angle  $\,\alpha_N\,$  between $\,\bfq_2\,$ and  $\,{\bf Q}_N\,$   is approximately equal to the angle between $\,\bfq_2\,$  and $\,\bfp_1$.  As a consequence, the solutions of Eq. (\ref{FN0}) have, with  a  very good approximation, a rotational symmetry with respect to the direction of the incident electron direction.

 We remark also that  because in the regime $\,(\gamma_1\gg\eta\sim1)\,$ the dressed momentum $\,{\bf q}_2\,$ is very close to $\,{\bf p}_2$, the two distributions, one  expressed  in terms of the variables  of $\,{\bf q}_2\,$, the other in terms of   $\,{\bf p}_2\,$,  are almost identical.

In order to obtain     the  energy  distribution or the angular distribution of the scattered electrons, one needs to integrate the  double differential distribution (\ref{distr}) over the parameters that are not observed,  by writing   the $\,\delta$-function in  a way  convenient for  each distribution.

 For the {\it angular distribution} we need the relation:
\begin{equation}\label{deW}
\delta(F_N({\bf q}_2))=\sum_{\mathrm{sol}}\frac{\delta(X-X^{\mathrm{sol}})}{\mid\frac{\partial F_N}{\partial X}\mid}=\frac{1}{m_*c}\,\sum_{\mathrm{sol}}\frac{q_2^0\,|{\bf Q}_N-{\bf q}_2|}{|{\bf Q}_N|\,\sqrt{\cos^2\alpha_N-C_N}}\,\delta(X-X^{\mathrm{sol}})\,,
\end{equation}
which is based on the solutions (\ref{solxi})  and (\ref{solxii}) of Eq. (\ref{FN0}) for the unknown  $\,X\equiv|{\bf q}_2|/(m_*c)$;  the quantity $\,C_{\mathrm N}\,$ is defined in (\ref{alN}). By the generic summation index '${\mathrm{sol}}$',  we understand  the   (one or two) solutions acceptable  at fixed $N$  (see details in Appendix A). When this expression is replaced in (\ref{distr}),  it displays  the   position of the singularities  in $\,X\,$ at fixed direction of $\,\bfq_2\,$. 

 For $\,\frac{\partial F_N}{\partial X}=0\,$ the expression in (\ref{deW})  for $\,\delta(F_N({\bf q}_2))\,$ is not valid. In the following,  we work with $\,\cos\alpha_N < \sqrt{C_N} $ and after obtaining the angular distributions, we  take the limit $\,\cos\alpha_N = \sqrt{C_N} $. As shown further, the   singularity present in the double differential distribution (\ref{distr})   influences  the angular distribution of the electrons, obtained after integration on the scattered electron energy. 

To prepare (\ref{distr}) for the calculation of the {\it energy  distribution},  we have  to find  the polar angles of $\,\bfq_2\,$ that are solutions of (\ref{FN0})  at fixed $\,\mid \bfq_2\mid\,$. The equation (\ref{solY})  gives  us, for any $\,X\,$ in the interval (\ref{domX}), the unique solution  for $\,\cos\alpha_N\,$, denoted by  $\,Y(X)\,$  (see Appendix A). From it we derive the possible values for  the polar angles of $\,{\bf q}_2\,$ by solving the equation
\begin{equation}\label{eqtp}
\cos\theta_N\cos\theta_{q_2}+\sin\theta_N\sin\theta_{q_2}\cos(\phi_{q_2}-\phi_N)=Y(X)\,,
\end{equation}
 considering as the unknown variables one of the two angle $\,\theta_{q_2}\,$ or $\,\phi_{q_2}\,$,  with fixed $\,Y(X)$.

Simple particular cases are  collinear and head-on collisions    [$\,\sin\theta_N=0\,$, $\,\cos\theta_N=\pm1\,$], when  Eq. (\ref{eqtp}) is an equation for $\,\theta_{q_2}$  only, with the solution  $\,\cos\theta_{q_2}=\sigma_N Y(X)\,$
with $\,\sigma_N={\mathrm{sgn}}(\cos\theta_N)$. In this case  we write the $\,\delta$-function in (\ref{distr})  as 
\begin{equation}\label{det}
\delta(F_N({\bf q}_2))=m_*c\,\frac{|{\bf Q}_N-{\bf q}_2|}{|{\bf q}_2|\,|{\bf Q}_N|}\delta(\cos\theta_{q_2}-\sigma_N Y(X))\viq \sin\theta_N=0\,.
\end{equation}

The cases $\,\theta_{N}\ne0,\pi\,$  are more complicated since both angles $\theta_{q_2}$ and $\phi_{q_2}$ appear in Eq. (\ref{eqtp}). One  possibility is to  solve Eq.  (\ref{eqtp}) for the unknown $\,\phi_{q_2}$, keeping as parameter $\,\theta_{q_2}$. As shown in Appendix B,  Eq. (\ref{eqtp})  has two solutions,
\begin{equation}
\phi^{\,\mathrm{sol}}=\phi_N\pm\Phi_0,\quad \Phi_0=\arccos\left[\frac{Y(X)-\cos\theta_N\cos\theta_{q_2}}
{\sin\theta_N\sin\theta_{q_2}}\right]\,.\label{phi-sol}
\end{equation}
if $\,\theta_{q_2}\,$ obeys the condition 
\beq\label{condte}
\cos\theta_{q_2}\in[\cos(\theta_N+\alpha_N),\cos(\theta_N-\alpha_N)]\,. 
\eeq
 This condition defines an angular range that we denote by $\,{\cal I}_\theta$. We  emphasize that in the present context the value taken by  the angle $\,\alpha_N$  depends on $|{\bf q}_2|$, being expressed as $\alpha_N=\arccos(Y(X))$.  Finally, the procedure leads to the expression of $\delta$-function in (\ref{distr})
\begin{equation}\label{defi}
\delta(F_N({\bf q}_2))=m_*c\,\sum\limits_{\mathrm{sol}}\frac{|{\bf Q}_N-{\bf q}_2|}{|{\bf q}_2||{\bf Q}_N|}\frac1{|\sin\theta_N\sin\theta_{q_2}\sin\Phi_0|}\,
\delta(\phi_{q_2}-\phi^{\,\mathrm{sol}})\viq \sin\theta_N\not=0\,.
\end{equation}
With this expression of $\,\delta(F_N)\,$ the distribution (\ref{distr}) displays the position of the singularities in $\,\phi_{q_2}\,$ at fixed 
$\,\theta_{q_2}\,$ and $\,\mid \bfq_2\mid $.

  If we choose to solve the equation (\ref{eqtp}) for the unknown $\,\theta_{q_2}$ with $\,\phi_{q_2}$ as a parameter,  the solutions are more complicated. With the notation $\, u_{q_2}=\cos\theta_{q_2},$ one finds (for details, see Appendix B)  two possible solutions: 
\begin{eqnarray}\label{solecuu}
u_{q_2}^{(\pm)}&=&\frac{1}{s_N}\,\left(Y(X)\cos\theta_N\pm\sin\theta_N|\cos(\phi_{q_2}-\phi_N)|\sqrt{s_N-Y^2(X)}\right).\\
&&s_N\equiv \cos^2\theta_N+\sin^2\theta_N\cos^2(\phi_{q_2}-\phi_N)\,.
\end{eqnarray}
Depending on  the initial conditions   and on  the value of $\,\phi_{q_2}$, one or both  solutions are acceptable, namely:

 i) for $\,Y^2(X)\le\cos^2\theta_N$,    only one solution is acceptable for any value of $\,\phi_{q_2}\,$:  $\,u_{q_2}^{(+)}$,  if $\cos\theta_N\cos(\phi_{q_2}-\phi_{N})<0$,  and $\,u_{q_2}^{(-)}$, if $\cos\theta_N\cos(\phi_{q_2}-\phi_{N})>0\,$,

ii) for $\,\cos^2\theta_N<Y^2(X)\le1$,  both solutions $u_{q_2}^{(\pm)}$, are acceptable, but the domain of  $\,\phi_{q_2}$ is reduced to $\,\phi_{q_2}\in[\,\phi_N-\phi_0,\phi_N+\phi_0\,]\,$ if $\,Y(X)>0\,$ and to $\,\phi_{q_2}\in[\,\pi+\phi_N-\phi_0,\pi+\phi_N+\phi_0\,]\,$ if $\,Y(X)<0\,$, where $\,\phi_0=\arccos\sqrt{(Y^2(X)-\cos^2\alpha_{N})/\sin^2\alpha_N}\,\in(0,\pi/2)\,$. 

\noindent
The  $\,\delta$-function in (\ref{distr}) is written now as
\begin{equation}
\delta(F_N(\mid\bfq_2\mid; u_{q_2},\phi_{q_2}))=m_*c\,\sum\limits_{\mathrm{sol}}
\frac{|{\bf Q}_N-{\bf q}_2|\sqrt{1-(u_{q_2}^{\mathrm{(sol)}})^2}}{2\,|{\bf Q}_N|\,|{\bf q}_2|}\,\frac{\delta\left(u_{q_2}-u_{q_2}^{(\mathrm{sol})}\right)}
{\sqrt{s_N-Y^2(X)}}\,.\label{tr-delta1}
\end{equation}
It gives  the position of the singularities in $\,\theta_{q_2}\,$ at fixed $\,\phi_{q_2}\,$ and $\,\mid \bfq_2\mid $.

\subsection{Angular distribution of electrons}\label{s-iiib}

  We get the angular distribution of electrons  using  the expression (\ref{deW}) in the distribution (\ref{distr}) and integrating over $|{\bf q}_2|=(m_*c)\,X$, with the result
\begin{equation}
\frac{d\Gamma_e}{d\Omega_{q_2}}= {\sum\limits_{N\ge 1} \sum_{\mathrm{sol}}\frac{|{\bf q}_2|^2\,W_2\,|{\bf Q}_N-{\bf q}_2|}{c\,|{\bf Q}_N|}\,
\frac{\Gamma^{(N)}({\bf q}_2)}{\sqrt{\cos^2\alpha_N-C_N}}\vline\,}_{|{\bf q}_2|=m_*cX^{\mathrm{sol}}}\,
\,\equiv\sum_N \Gamma_{e,ang}^{(N)}.\label{rataang}
\end{equation}
The substitution rule indicated above means that  the modulus of $\,{\bf q}_2\,$ must be replaced everywhere by  $\,(m_*c)X^{\mathrm{sol}}$, with $\,X^{\mathrm{sol}}\,$ given by Eq. (\ref{solxi}) or (\ref{solxii}).

Based on  the results in Sect. \ref{s-iiia}, Appendix \ref{s-a} and some more details given in  Appendix \ref{s-b},  we mention   here the main features of the angular distribution. We describe the situation of a term $\,\Gamma_{e,ang}^{(N)}$, with a fixed value of $\,N$.   If, for  that $\,N,$ we are in the case I, when  $\,C_N<0,$ [$\,C_N\,$ defined in (\ref{alN})] there is one solution $\,X^{\mathrm{sol}}\,$ [Eq. (\ref{solxi})] for any $\,\cos\alpha_N \in[0,\pi]$, i.e. for any direction of the scattered electron,  and the  term $\,\Gamma_{e,ang}^{(N)}\,$ in the sum (\ref{rataang}) is finite.  If,  for the considered $\, N \,$, we are in the case II, when  $\,0<C_N<1$, then, there are two solutions $\,X^{\mathrm{sol}}\,$ [Eq. (\ref{solxii})] for any direction $\,\alpha_N\,$ obeying the condition $\,\cos\alpha_N\le\sqrt{C_N}$.  For $\,\cos\alpha_N=\sqrt{C_N}$, the two solutions $\,X_{\pm}\,$ in (\ref{solxii}) coalesce and the corresponding $\,\Gamma_{e,ang}^{(N)}\,$ has a singularity.   The condition  $\,\cos\alpha_N\le\sqrt{C_N}$, determining the possible scattering angles for a given $\,N$   in the case II, can be expressed in terms of polar angles of the electron  in the form $\,\{\theta_{q_2},\phi_{q_2}\}\in{\cal D}(C_N)$. The explicit expression of the domain $\,{\cal D}(C_N)\,$ is deduced in Appendix B.
 
\subsection{Energy distribution of electrons}\label{s-iiic}

The energy distribution is obtained  by integrating the fully differential distribution (\ref{distr}) over the electron directions determined by the angles $\,\theta_{q_2}$ and  $\,\phi_{q_2}$.

 For  collinear and head-on collisions, using (\ref{det}), the integral over $\,\theta_{q_2}\,$ is performed directly and the energy distribution becomes 
\begin{equation}
\frac{d\Gamma_e}{dW_2}=\frac{m_*}{c}\,\sum_{N\ge1}\int\limits_0^{2\pi}d\phi_{q_2}
{\frac{W_2|{\bf Q}_N-{\bf q}_2|}{|{\bf Q}_N|}\,\Gamma^{(N)}({\bf q}_2)\vline}_{\,\theta_{q_2}=\theta_{\mathrm{sol}}}=\sum_{N\ge1}\Gamma_{e,W}^{(N)}\viq \sin\theta_N=0\,,
\end{equation}
where  $\,\theta_{\mathrm{sol}}=\arccos(Y(|{\bf q}_2|/(m_*c)))\,$, if $\sigma_N=1$, and   $\,\theta_{\mathrm{sol}}=\pi-\arccos(Y(|{\bf q}_2|/(m_*c)))$ if $\sigma_N=-1\,$,  with $Y(|{\bf q}_2|/(m_*c))$ calculated according to (\ref{solY}).

For   other  initial configurations,  it is convenient to use in (\ref{distr}) the expression  (\ref{defi})  of the $\,\delta$-function and,  as a consequence, in the calculation of the energy distribution the integral over $\,\phi_{q_2}$ is performed directly. After that, for the integral on $\,\theta_{q_2}\,$ that has to be done numerically, the domain of integration  reduces to the interval   $\, {\cal I}_{\theta}$  defined by the condition  (\ref{condte}).  The final result reads:
\begin{equation}
\frac{d\Gamma_e}{dW_2}=\frac{m_*}{c}\,\sum_{N\ge1}\sum\limits_{\mathrm{sol}}\int\limits_{{\cal I}_{\theta}}\frac{d\theta_{q_2}}{\sin\theta_N\sin\Phi_0}{\frac{W_2|{\bf Q}_N-{\bf q}_2|}{|{\bf Q}_N|}\Gamma^{(N)}({\bf q}_2)\vline}_{\,\phi_{q_2}=\phi^{\,\mathrm{sol}}}=\sum_{N\ge1}\Gamma_{e,W}^{(N)}
\end{equation}
 with $\,\phi^{\,\mathrm{sol}}\,$ and $\,\Phi_0\,$ given by (\ref{phi-sol}).

\subsection{An example}\label{s-iiid}

 We illustrate the previous analysis by an example. We have seen that the $\,\delta$-function in the multiple differential distribution (\ref{multi}) imposes the restriction (\ref{FN0}) on the vector $\,\bfq_2\,$  and we have described the position  of the singularities in terms of  ($\,W_2,\,\theta_{q_2},\,\phi_{q_2}\,$).   As mentioned   in  Sect. \ref{s-iiia} the analysis can be converted in terms of the bare momentum $\,\bfp_2\,$. In this case we think of  surfaces in the space ($\,E_2,\,\theta_{p_2},\,\phi_{p_2})\,$  on which the singularities are localized. In the example that follows we shall  present graphs with the  curves giving the position of the singularities in the plane ($\,E_2,\,\theta_{p_2}$) at fixed $\,\phi_{p_2}=\phi_{p_1}$.

  We choose the case of an electron of  energy $\,E_1=46.6$ GeV scattered by a circularly polarized  monochromatic wave with the frequency $\,\omega_1=0.043$ a.u. (1.17 eV) and the field intensity $\,I=4.4\times 10^{17}$ W/cm$^2$ ($\eta=0.6$); these conditions are close to those in the SLAC experiment, in which the  detection of NLCS  was achieved.    We consider two cases for the initial direction of the electron: (a) $\theta_{p_1}=0.9\,\pi$, close to  the value used at SLAC,  and (b) $\,\theta_{p_1}=0.5\,\pi$  (orthogonal geometry),  with  $\,\phi_{p_1}=0\,$ in both cases.  For these initial conditions we have $n_1\cdot q_1>m_*c$ and we are in the case II (defined in III.A) for values of $\,N\,$ up to $\approx10^6$; this value is much larger  than the maximum value of $\,N\,$ contributing to the electron distribution  at the intensity considered,  which is of the order of ten.

We describe the position of the $\,\delta$-type singularities in the double differential distribution (\ref{distr})  based on Eq. (\ref{tr-delta1}), giving some   details valid in our particular case,  $\,\phi_{p_1}=\phi_{p_2}=0\,$, using as  variables  the  bare  energy $\,E_{p_2}\,$ and   polar angle $\,\theta_{p_2}$.   In the present discussion, preceding Fig. 1, we have in mind  only low values of $N$ (of the order of ten), for which, as we have mentioned before, we are in the case I. In the particular case $\phi_{p_2}=\phi_{p_1}=0$  we have chosen, for any $\,X\,$ in the interval (\ref{domX}) (i.e. for any energy $\,W_2$ in (\ref{Wab})) the two acceptable solutions of Eq.  (\ref{eqtp}), given by Eq. (\ref{solecuu}),  reduce to $\,\theta_{q_2}=\theta_N\pm\arccos Y(X)\,$ which coalesce for $\,Y(X)=1$, i.e. at the ends $\,W_2=W_a\,$ and $\,W_2=W_b\,$ of the interval. 
The maximum domain of variation for the angles is given by the condition
\begin{equation}
\theta_{q_2}\in[\,\theta_{q,A}=\theta_N-\arccos Y_{\mathrm{min}},\theta_{q,B}=\theta_N+\arccos Y_{\mathrm{min}}\,]\,,
\end{equation}
with $\,Y_{\mathrm{min}}\,$ given by (\ref{Ymin}).
This  domain can be transcribed in terms of bare energy and scattering angle, using the relation
\begin{equation}
\theta_{qA,B}\rightarrow\theta_{pA,B}=\arccos\left(\frac{|{\bf q}_2|\cos\theta_{qA,B}-m^2c^2\eta^2/(4n_1\cdot q_2)}{\sqrt{(W_2-m^2c^3\eta^2/(4n_1\cdot q_2))^2/c^2-m^2c^2}}\right)\label{thetapAB}
\end{equation}
for the angles and Eq. (\ref{EW}) for the energy. 

In the regime discussed here ($\,\gamma\gg\eta\sim1$\,), $\,{\bf q}_1\,$ and $\,{\bf q}_2\,$ are very close to the corresponding bare momenta $\,{\bf p}_1\,$ and respectively $\,{\bf p}_2$,  so the results in the plane $\,(E_{p_2},\theta_{p_2})\,$ are practically identical at the graphical level to those in $\,(W_2,\theta_{q_2})$;  in particular, we have $W_{a,b}\approx E_{a,b}$ and $\theta_{pA,B}\approx\theta_{qA,B}$. Another particularity is that the upper limits of the intervals $\,W_b(N)\approx E_b(N)$, defined in (\ref{Wab}), are almost independent of $\,N\,$ and approximately equal to the initial electron energy $\,E_1$; the lower limits, however, are significantly dependent of $\,N$.  Then, the energy of the final electron in the process in which $N$ photons are absorbed takes values in an interval $E_2\in(E_a(N),E_1)$ with $E_a(N)<E_a(N-1)$. The successive  values $E_a(N)$ are named {\it thresholds} of the energy spectrum.

\begin{figure}
\includegraphics[scale=0.18]{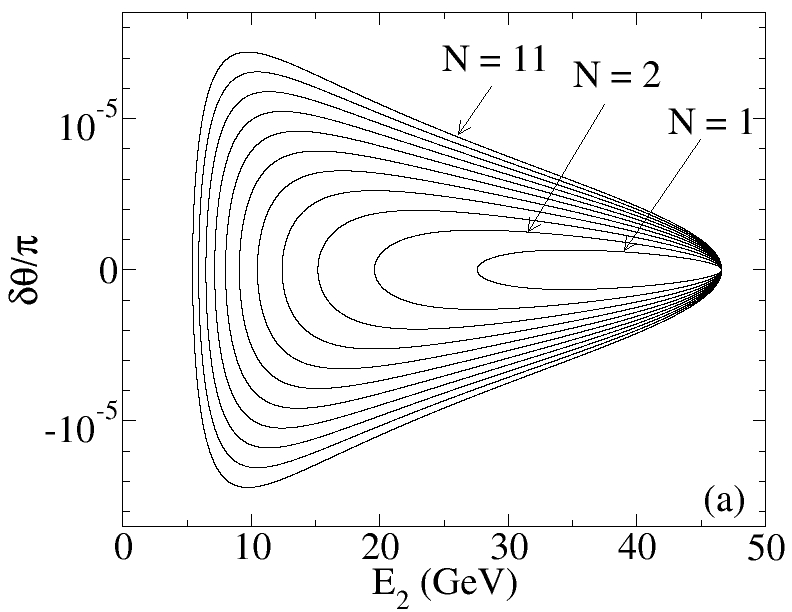}
\includegraphics[scale=0.18]{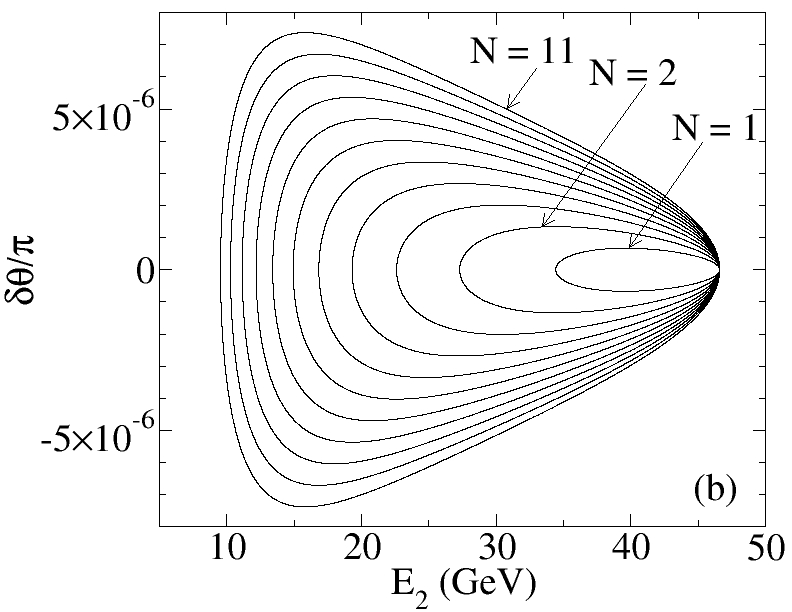}
\includegraphics[scale=0.18]{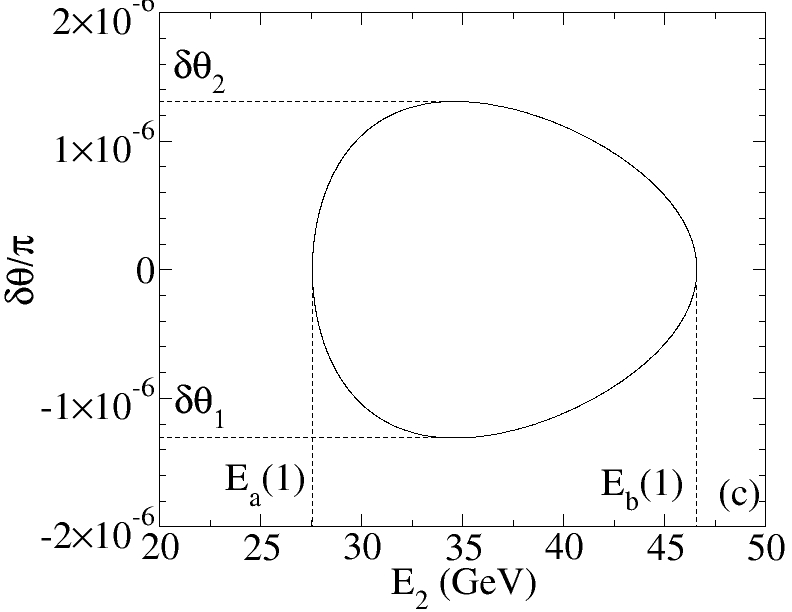}
\caption{Position of the first eleven lines in the plane $(E_2,\theta_{2})$ for $E_1=46.6$ GeV, $\eta=0.6$,  circular polarization and two directions of the incident electron: (a) $\theta_{p_1}=0.9\pi$ and (b) $\theta_{p_1}=0.5\pi$; (c): detailed view of (a) for $N=1$\,.\label{fig-mono}}
\end{figure}

 In  Figures \ref{fig-mono} (a) and (b), 
 for  the two  values of $\,\theta_{p_1}\,$ we have chosen for the direction of the incident electron, we display  in the plane $\,(E_{p_2},\theta_{p_2})\,$ the curves that represent the solutions of Eq. (\ref{FN0}) for   $\,N$ taking values from 1 to 11. 
   The coordinate along the $\,y$ axis  is $\,\delta\theta\,$,  defined as $\,\delta \theta\equiv\theta_{p_{2}}-\theta_{p_{1}}$; the good symmetry of the two figures with respect to the value $\,\delta\theta=0\,$ is a consequence of the rotational symmetry around the direction of $\,\bfp_1\,$  mentioned  in Sect. \ref{s-iiia}. The results show that $\,\delta\theta\,$ takes  very small values for all $\,N$, i.e. the final electron direction is very close to the initial one for all the  cases represented ($\,N\le11$).  On  the other hand,  the energies $\,E_2\,$ are  spread till relatively small values. The interpretation of these results is that in  case (a) the electron can  lose up to $45\%$ of its initial energy in the process in which only one photon is absorbed ($\,N=1$), and up to $85\%$ for $\,N=11$. In  case  (b), when  the initial electron incident orthogonal on the laser pulse  direction,   the angular distribution is more compressed towards small angles  and it is also compressed in the region of larger final energies. As, according to the conservation laws, the energy lost by the electron is converted in the energy of the emitted photon,  this means that  this case is less efficient for energy conversion.

In Fig. \ref{fig-mono} (c) is presented   only the curve with $\,N=1$ from the case (a). The limits $\,E_a(1\,)$ and  $\,E_b(1)\,$  of the domain in energy that gives contribution to the spectrum,  obtained from   Eq. (\ref{Wab}) using Eq. (\ref{EW}),  are marked on the graph; for any  $\,E_2\in \left(E_a(1),E_b(1)\right)\,$ there are two  angles $\theta_{q_2}$ for which the equation (\ref{FN0}) is verified, they become a double root for  $\,E_2=\,E_a(1)$ and  $\,E_2= E_b(1)$. As discussed after Eq. (\ref{thetapAB})  the value $\,E_a(1)\,$ corresponds to the threshold of one photon absorption in the energy spectrum. 
The domain of angles $\theta_{p_2}$ that contribute to the angular distribution  for $\,N=1\,$ is the interval $\,(\theta_{p_1}+\delta\theta_1,\theta_{p_1}+\delta\theta_2)$; within this interval there are two solutions $E_2$ of Eq. (\ref{FN0}), which coalesce for $\,\theta_{p_2}=\theta_{p_1}+\delta\theta_1\,$ or $\,\theta_{p_2}=\theta_{p_1}+\delta\theta_2$.

 Another aspect  worth  to be discussed  is   which would be  the dependence of the type of curves represented in Fig. \ref{fig-mono} (a) and (b) on the laser intensity.  One feature to be considered is the increase of the maximum number of photons that can be absorbed in the process  with the laser intensity.   The other feature is that, at fixed $\,N\,$ and $\,q_1$, when the laser intensity increases the curves tend to become closer to each other, i.e. the domain in which the energy of the final electron can take values shrinks. This process  can be  understood  using the concept of ``dressed mass'': when the laser intensity increases the electron becomes heavier, and consequently its recoil at fixed number of photons absorbed decreases.

\section{The plane wave pulse} \label{s-iv}

We consider now the more realistic model of a pulsed plane wave, going back to Eqs. (\ref{s4}) and (\ref{Pi4}).  In the pulse case, the integral $\,{\cal B}\,$ is expressed  in terms of convergent integrals, using  Eq.(30) of  \cite{BF}   (see also, \cite{AI} for an alternative approach). 

 Now, as only three $\,\delta$-functions appear in (\ref{s4}), only three conditions  are imposed to  the six variables $\,\bfp_2\,$ and $\,\bfq_2$, namely
\beq
{\bf p}_{1\perp}-{\bf p}_{2\perp}-\hbar {\bf k}_{2\perp}=0\viq n_1\cdot( p_1-p_2-\hbar k_2)=0\,.\label{delta}
\eeq
In order to get the one-particle (electron of photon) distribution, the   differential distribution (\ref{s4}) is integrated over the momentum of the  that is not detected using the conservation rules (\ref{delta}). The integration over the orthogonal components of the  momenta is performed directly using the  $\,\delta$ function,  so from  (\ref{delta}) we get the replacement rules  to be used in (\ref{Pi4}),
\begin{equation}
{\bf p}_{2\perp}\rightarrow\widetilde{\bf p}_{2\perp}={\bf p}_{1\perp}-\hbar{\bf k}_{2\perp}\label{cons1}
\end{equation}
for the integration over the orthogonal component of the {\it photon} momentum, and respectively,
\begin{equation}
\hbar{\bf k}_{2\perp}\rightarrow\hbar\widetilde{\bf k}_{2\perp}={\bf p}_{1\perp}-{\bf p}_{2\perp}\label{cons2}
\end{equation}
for the integration over the orthogonal component of the {\it electron} momentum. 
The integration over the third component requires some further calculation, due to the fact that the  second relation in (\ref{delta})  contains a combination of energy and momenta.  We present the results in both cases. Using the adequate   relations from the following ones,
\beq
\delta[ n\cdot( p_1-p_2-\hbar k_2)]= \frac{\widetilde E_2}{cn_1\cdot (p_1-\hbar k_2)}\,\delta(p_{2z}-\widetilde p_{2z})=\frac{\hbar\widetilde\omega_2}{cn_1\cdot (p_1-p_2)}\,\delta(k_{2z}-\widetilde k_{2z})\,,
\eeq
with 
\beq
\widetilde p_{2z}=\frac{(mc)^2+({\bf p}_{1\perp}-\hbar{\bf k}_{2\perp})^2}{2n_1\cdot(p_1-\hbar k_2)}-\frac{n_1\cdot(p_1-\hbar k_2)}2,\quad \hbar\widetilde k_{2z}=\frac{({\bf p}_{1\perp}-{\bf p}_{2\perp})^2}{2n_1\cdot(p_1-p_2)}-\frac{n_1\cdot(p_1-p_2)}2,\label{cons3}
\eeq
and
\beq
\frac{\widetilde E_2}c=\frac{(mc)^2+({\bf p}_{1\perp}-\hbar{\bf k}_{2\perp})^2}{2n_1\cdot(p_1-\hbar k_2)}+\frac{n_1\cdot(p_1-\hbar k_2)}2,\quad \frac{\hbar\widetilde \omega_2}c=\frac{({\bf p}_{1\perp}-{\bf p}_{2\perp})^2}{2n_1\cdot(p_1-p_2)}+\frac{n_1\cdot(p_1-p_2)}2\,,\label{cons4}
\eeq
 one obtains the two   one-particle (photon or electron) distributions,
\beqa
d^2\Gamma_\gamma&=&\frac{\widetilde E_2}{cn_1\cdot \widetilde p_2}\,\frac{\omega_2^2}{c^3}\,\Pi_4(k_2,\widetilde p_2)\,d\omega_{2}d\Omega_{k_2} \,,\label{dist-ph}\\
d^2\Gamma_e&=&\frac{E_2}{cn_1\cdot\hbar \widetilde k_2}\,\frac{\widetilde \omega_2|{\bf p}_2|}{\hbar^2 c^2}\,\Pi_4(\widetilde k_2,p_2)\,dE_2\,d\Omega_{p_2} \,.\label{dist-el}
\eeqa
The attribute {\it unpolarized} was omitted.

 NB. The quantity denoted here by $\,\tilde k_2\,$ is different  from that defined in   (\ref{cons}) and used in Sect. \ref{s-iii}, as it comes out from a different conservation rule.

The  structure of the previous two distributions, using each in a specific way  the same function $\,\Pi_4\,$  implies the   possibility of  connections between the two distributions,  as it will be mentioned  at the end of  Sect. \ref{s-vb}.

\section{Numerical results}\label{s-v}

 We  consider  two type of pulses: i) a  pulse with a finite duration,  of almost rectangular  shape,  with the envelope $\,f(\phi)\,$ in Eq. (\ref{puls})  constant on a region of length equal to a multiple $N_c$ of periods of the carrier, and two very short smooth wings; we shall name this pulse ``quasimonochromatic'', ii) a  pulse without  a constant region, consisting in two wings of variable length. We
 shall see that the first type of pulse,  if $\,N_c$ is large enough,   leads to results similar to those predicted by the monochromatic approximation, which explain the adopted terminology.  The contribution  to the scattering probability of the wings of the pulse is very small compared to the contribution of the flat central region, still  these smooth wings are required in order to ensure the continuity of the vector potential and of its derivative. The results obtained with the second pulse  are considerably different from the monochromatic ones.

\begin{figure}
\includegraphics[scale=0.23]{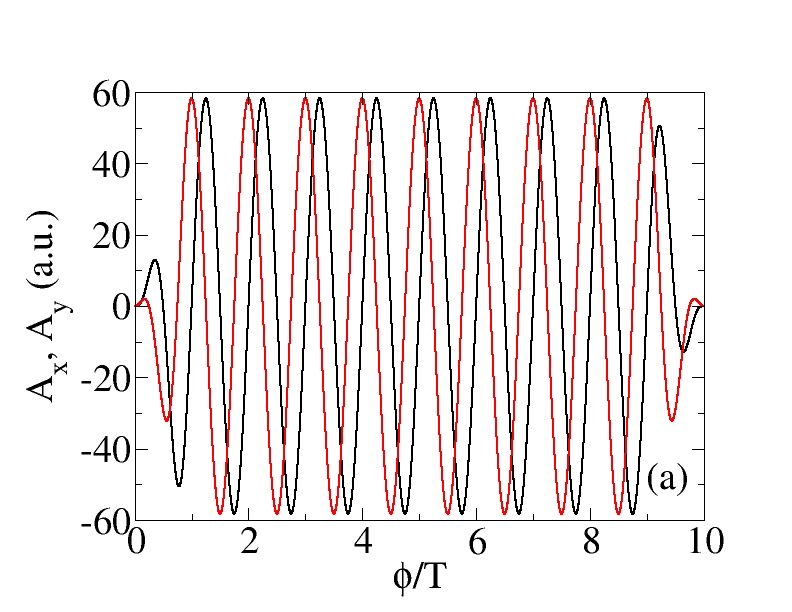}
\includegraphics[scale=0.23]{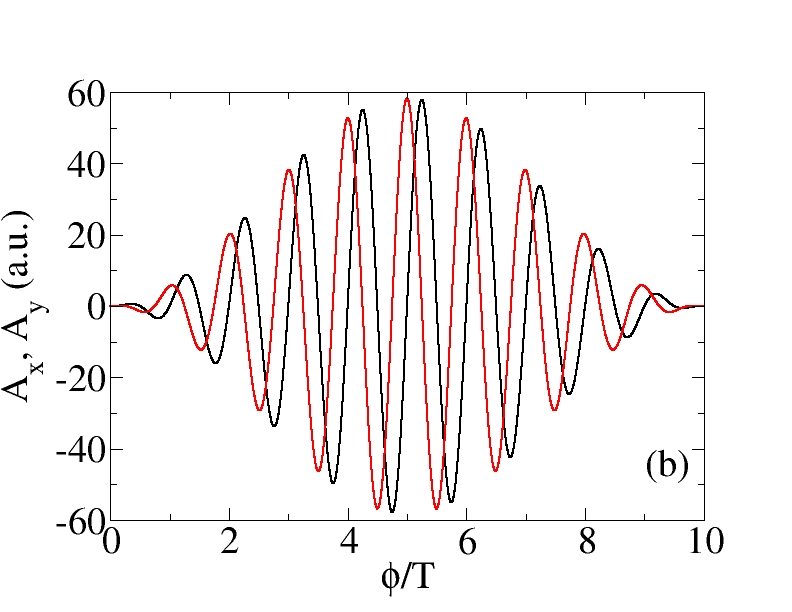}
\caption{(Color online) The components $\, A_x\, $ and  $\, A_y\, $ of the vector potential  for a rectangular pulse with $N_c=8$ cycles (a) and for a $\cos^2$ pulse with $N_t=10$ cycles.\label{fig-pulse}}
\end{figure}

In Fig. \ref{fig-pulse} is represented the vector potential $\,{\bf A}\,$ as a function on $\,\phi/T\,$ for the two pulses  mentioned before. The rectangular portion of the pulse in   (a)  has the length   of  $\,N_c$=8  periods of the carrier.  In the  case (b) we have chosen a $\,\cos^2$ envelope, the total length of the pulse  corresponds to a number  $\,N_t$ of periods of the carrier equal to 10.  In both cases the parameter $\,\eta\,$ is 0.6 ( the value used in the SLAC experiment). In all numerical examples presented here we choose the laser central frequency $\,\omega_1=0.043$ a.u. (1.17 eV)  and circular polarization. 

\subsection{Effect of the pulse shape} \label{s-va}

We have calculated the electron double  differential probability distribution $\,d^2\Gamma_{\mathrm e}/dE_2d\Omega_{p_2}\,$ for the conditions of the SLAC experiment ($\,\omega_1=0.043$ a.u., $\,\eta=0.6$, $\,E_1=46.6$ GeV) and for  the two pulses represented in Fig. \ref{fig-pulse}. 

  In Fig. \ref{fig-slac3d} (a) and (b), we present results  in a logarithmic color scale in the same coordinates as those in Fig. \ref{fig-mono} $\,(E_2,\,\delta\theta=\theta_{p_2}-\theta_{p_1})$. The case considered is $\phi_{p_2}=\phi_{p_1}=0$, as in  Fig. \ref{fig-mono}.    Due to the very good symmetry of the results with respect to $\,\delta\theta=0$, remarked also in the discussion of the monochromatic case (Sect \ref{s-iiid}), only the values $\,\delta\theta<0$ are presented.   In both figures one can see a series of maxima   located  on  curves  with the same shape as  those presented in Fig. \ref{fig-mono} in the monochromatic case. Notice that in Fig. 1 both $\,\delta \theta<0$ and $\,\delta \theta>\,$ are represented.
\begin{figure}
\includegraphics[scale=0.3]{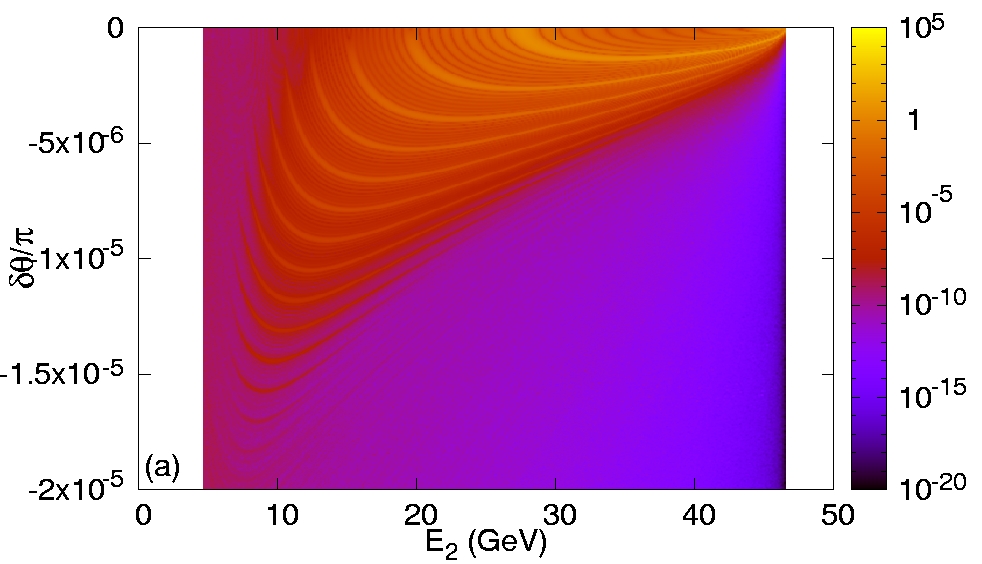}
\includegraphics[scale=0.3]{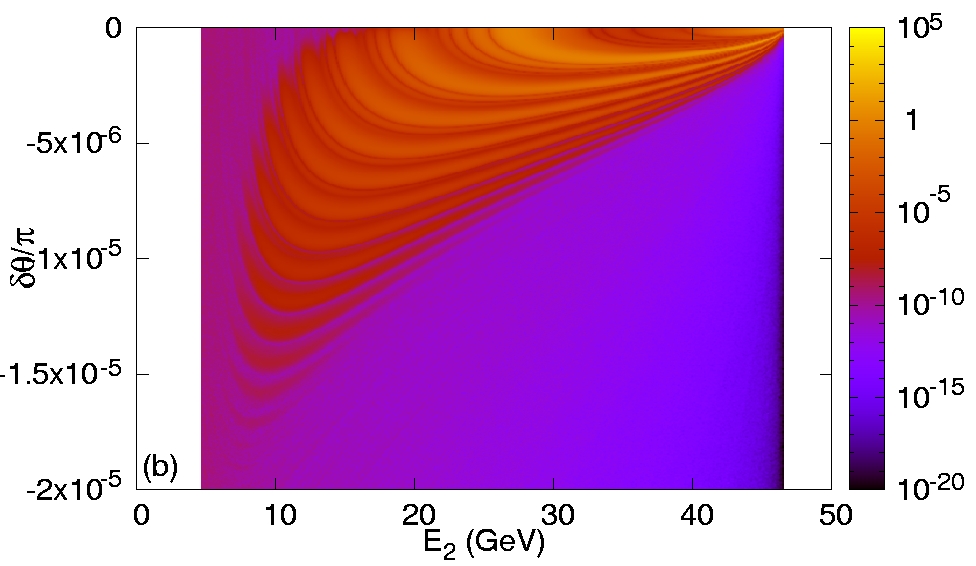}
\includegraphics[scale=0.2]{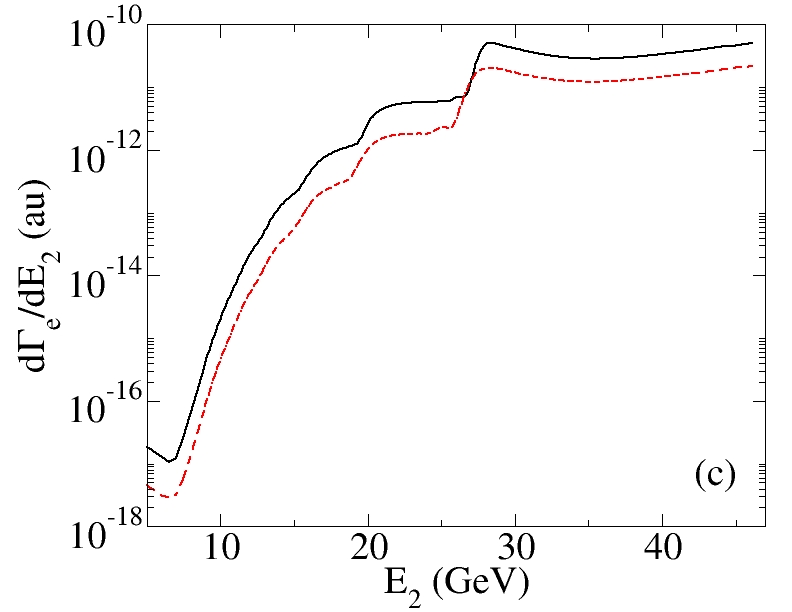}\hspace*{2cm}
\includegraphics[scale=0.2]{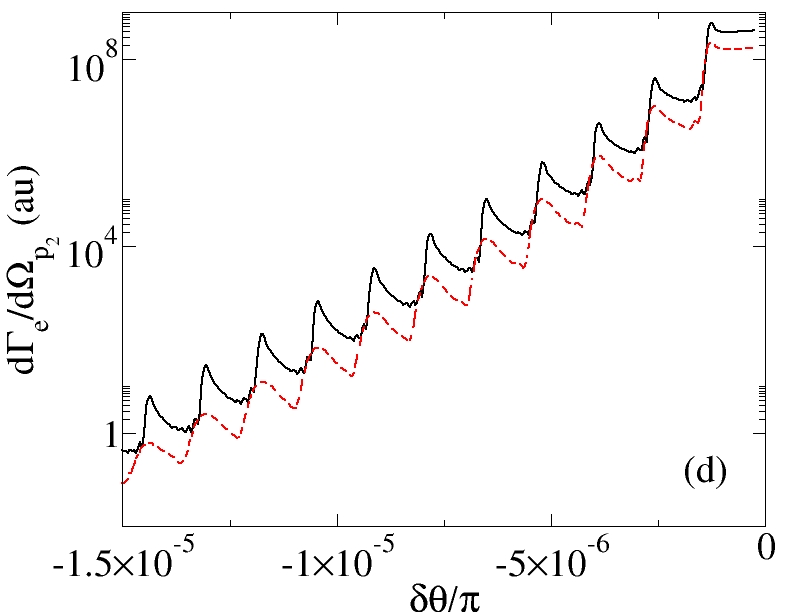}
\caption{(Color online) (a): Electron probability distribution $\frac{d^2\Gamma_e}{dE_2d\Omega_{p_2}}$ for the conditions of SLAC experiment and for a rectangular pulse; (b): the same as in (a), but for a $\cos^2$ pulse; (c): energy probability distribution for   the two pulses; (d): angular probability distribution for the two pulses. Black (full) lines refer to the case of rectangular pulse, and red (dashed) lines are for the $\cos^2$ pulse.\label{fig-slac3d}}
\end{figure}
For the  intensity  we consider,  with a relatively low value,  only  the first 11 maxima are visible. The difference between the two cases is that, while for the rectangular pulse the maxima are very sharp, and have a fine substructure, for the $\,\cos^2$ pulse  the main  maxima and their subpeaks  become wider and smooth. 

In Fig. \ref{fig-slac3d} (c) are represented   the energy distributions $\,d\Gamma_{\mathrm e}/dE_2\,$,  for the rectangular pulse (full black line) and  for the $\cos^2$ pulse  (dashed red line). The   two distributions are similar, having a ``ladder-like'' structure, with successive shoulders which can be understood  based on the monochromatic limit: their positions coincide  with the  thresholds $\, E_a(N)\approx W_a(N)\,$ in Eq. (\ref{Wab});  the upper limits $\,E_b(N)\,$ are almost independent of $\,N$ and approximately equal to  the initial electron energy $\,E_1$.  The first   interval at the right in Fig.  \ref{fig-slac3d} (c) covers the region $\,E_2\in(E_a(1)\approx27.6\,{\mathrm{GeV}},E_1)\,$ and can be interpreted as the sum of contributions of the processes in which any number of photons $\,N\ge1\,$ can be absorbed. The next step,  the region $\,E_2\in(E_a(2)\approx18.\,{\mathrm{GeV}},E_a(1)\approx27.6\,{\mathrm{GeV}})$, is the contribution of the processes with $\,N\ge2$,  as $\,N=1\,$ does not contribute anymore and so on. The fact that the values of the successive steps decrease very fast (note the logarithmic scale) is due to the  relatively small value of $\,\eta$, still close to the perturbative regime. The figure \ref{fig-slac3d} (c) is similar to  Fig. 4  in \cite{E144-2}, calculated  there in the monochromatic approximation.
 
The angular distribution $\,d\Gamma_{\mathrm e}/d\Omega_{p_2}\,$  for $\,\phi_{p_2}=\phi_{p_1}=0\,$ and variable $\,\theta_{p_2}\,$ for the same  two pulses  as before is presented in Fig. \ref{fig-slac3d} (d). Here one can see again the same ``ladder-like'' structure, but, unlike in the case of the energy distribution, there is a sharp maximum at the left end of each step. These maxima are the corespondent of the singularities of the angular distribution existent in the monochromatic case [see Sect. \ref{s-iiic} ]  and are localized at $\,\delta\theta_{pA}(N)$  given by (\ref{thetapAB}),  as presented in the example III.D. As expected,  they are much better defined for the rectangular pulse than  for  the $\,\cos^2$ pulse.

\subsection{Effect of the field intensity}\label{s-vb}

We  illustrate now  the influence of the laser intensity on the double differential distribution (\ref{dist-el})  of the electron. 

\begin{figure}
\includegraphics[scale=0.25]{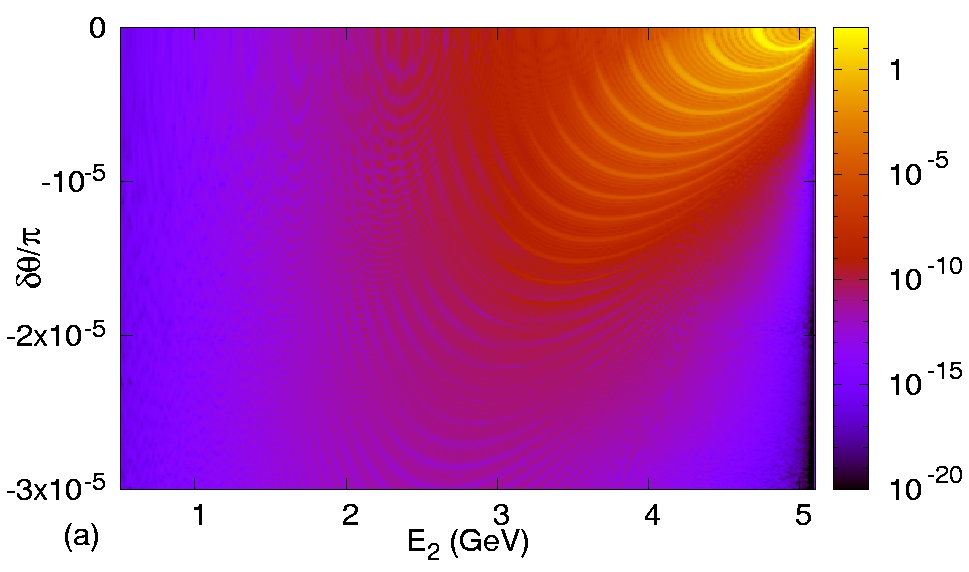}
\includegraphics[scale=0.25]{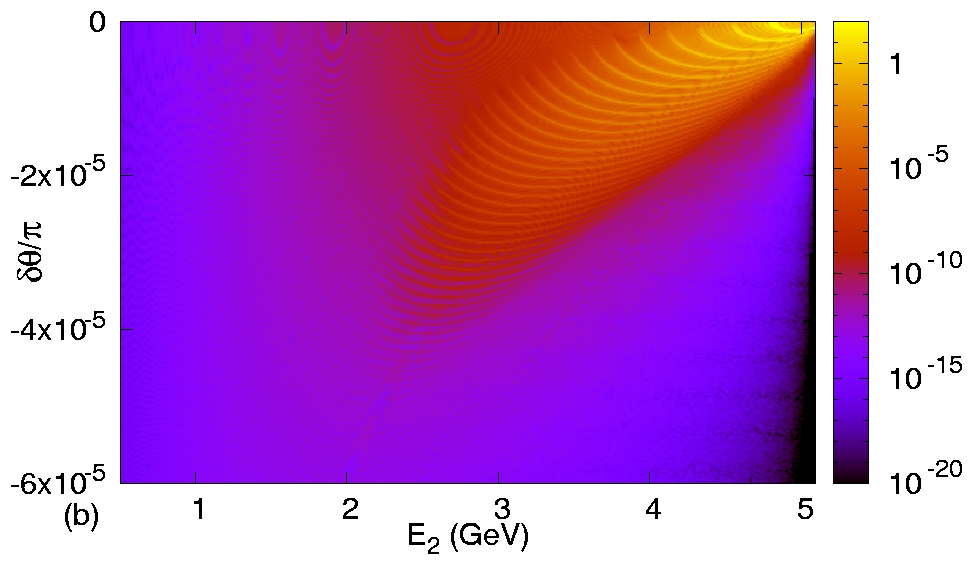}
\includegraphics[scale=0.25]{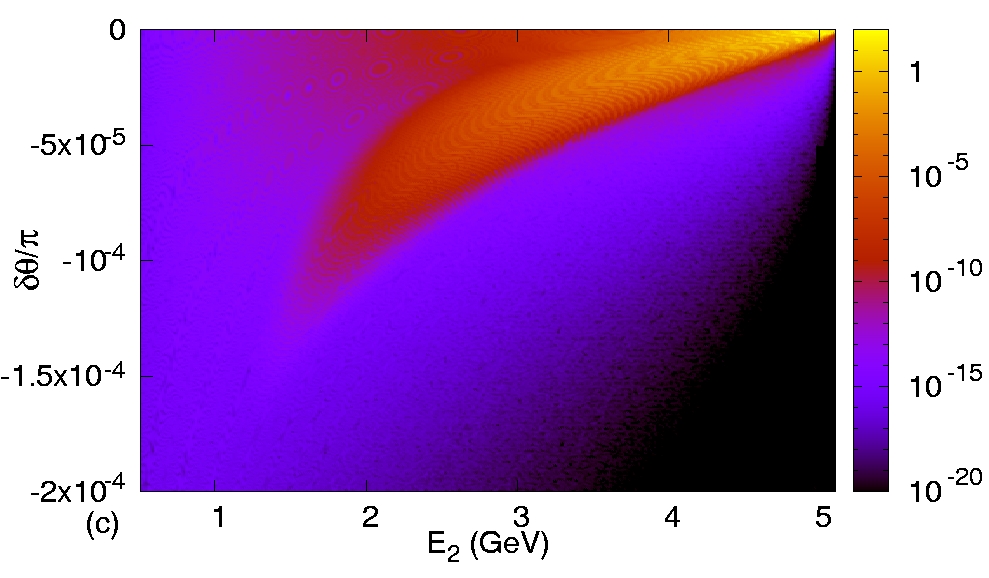}
\includegraphics[scale=0.25]{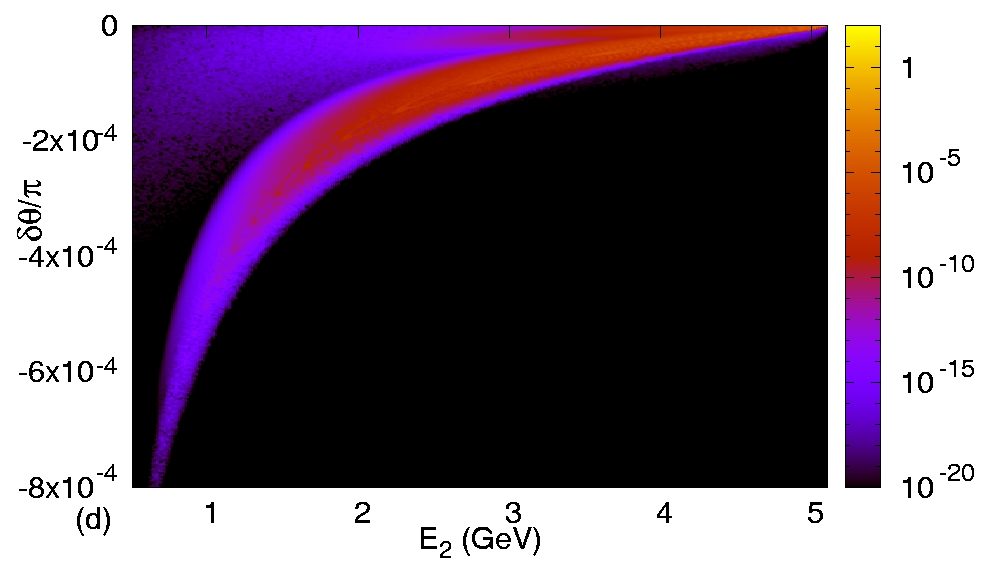}
\caption{(Color online) Electron probability distribution $\frac{d^2\Gamma_e}{dE_2d\Omega_{p_2}}$ as a function of $\gamma_2$ and $\delta\theta$ for head-on collision, initial electron energy $E_1=10^4\,mc^2$, and four values of the field intensity: $\eta=0.5$ (a), $\eta=1$ (b), $\eta=2$ (c), $\eta=5$ (d).\label{fig-int}} 
\end{figure}

 In Fig. \ref{fig-int} we consider  the case of a rectangular pulse with $\,N_c=6\,T$ and a  head-on collision ($\,\theta_{p_1}=\pi\,$) with the electron having the initial Lorentz factor $\,\gamma_1=10^4$. As in the previous subsection, we represent the probability distribution $\,\frac{d^2\Gamma_e}{dE_2d\Omega_{p_2}}\,$ in the plane $(\gamma_2,\, \delta\theta=\theta_{p_2}-\theta_{p_1})$.  We remind  that since we are in the case of head-on collisions and the laser is circularly polarized, the electron spectrum is symmetric with respect to rotations around the $\,z$ axis, i.e. it does not depend on the angle $\phi_{p_2}$. We have chosen four values of the parameter $\,\eta$: 0.5, 1, 2 and 4. For the first two values the spectra  present  a series of maxima  localized along curves whose shape and distribution is that  of the lines predicted in the monochromatic case (see  Fig. \ref{fig-mono}). However, when $\,\eta$ increases, the successive peaks become so close to each other that they start to overlap, tending to form a smooth continuum,  so we hardly distinguish them in  Fig.\ref{fig-int} (b) and not at all in  Figs. \ref{fig-int}(c) and (d).  This behaviour is in agreement with the discussion at the end of Sect. \ref{s-iii}. In the last two cases we remark   another interesting feature: for $\eta\ge2$ the distribution does nor cover uniformly the plane $\,( \gamma_2, \delta\theta)\,$ but  only a small region, with a well defined shape.

The  behaviour found in  Figs. \ref{fig-int} (c) and (d)  can be understood  in correlation with the {\it photon distribution}.  In  \cite{PhysSc} it was shown that in the case of an ultrarelativistic electron and for large  values of $\,\eta$,  the photons are emitted only in a well defined,  very small domain of angles;  although in the cited paper only the CED formalism is used, it can be shown that the conclusion concerning the photon distribution are valid also in the quantum case. For a rectangular pulse, as that considered here, and for head-on collisions, the emitted radiation has a continuous spectrum, extended from $\,\omega_2=0$ and up to a maximum value $\,\Omega_M$,  and it  is emitted  practically at a constant angle  $\,\theta_{0}$, symmetrically around the $z$ axis,   $\,\theta_{k_2}\in(\theta_0-\delta,\theta_0+\delta)$, $\,\delta\ll1$. In terms of  photon momentum, this means that the function  $\,\Pi_4(k_2,\widetilde p_2)$ in (\ref{s4})  is  non-negligible  only for $\,\bfk_2$ along the directions of the unit vectors 
$\,{\bf e}_0\equiv(\sin\theta_0\cos\phi,\sin\theta_0\sin\phi,\cos\theta_0)$, with  $\,\phi\in(0,2\pi)$; when expressed in terms of electron momentum, according to the conservation rules (\ref{cons2}), (\ref{cons4}), this condition  leads to the  particular shapes present in Fig. \ref{fig-int}  (c) and (d).   The correlation between the electron and photon distributions needs further investigation.

\subsection{Effect of the initial electron energy}\label{s-vc}

We present in Fig. \ref{fig-ef-gamma} the electron energy distribution $\,d\Gamma/dE_2\,$ for the case of a head-on collision, $\,\eta=5$, and for four values of the initial electron energy:  $\,\gamma_1=10$ ($E_1=5.11$ MeV) in full  line, $\,\gamma_1=10^2$ ($E_1=51.1$ MeV) in  dashed line, $\,\gamma_1=10^3$ ($\,E_1=511$ MeV) in  dotted line, $\,\gamma_1=10^4$ ($E_1=5.11$ Gev) in dash-dotted line.  The laser pulse is rectangular, with $\,N_c=10$ cycles. The values of the classicality parameter (\ref{classic1}) in the four cases are, respectively, $\,y=1.1\times10^{-4},\,1.1\times10^{-3},\,1.1\times10^{-2},\,1.1\times10^{-1} $. 

\begin{figure}
\includegraphics[scale=0.23]{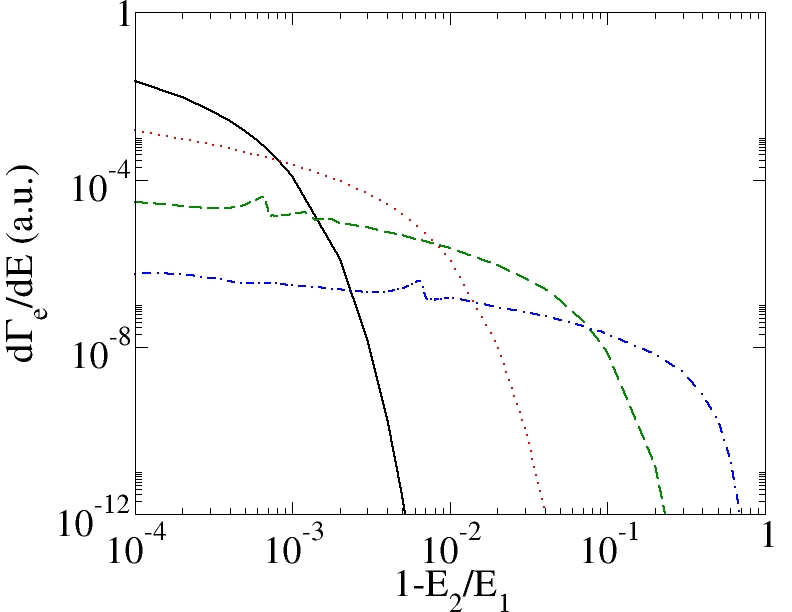}
\caption{(Color online) The energy distribution $d\Gamma_e/dE_2$ for $\eta=5$ and four values of the initial electron energy: $E_1=5.1$ MeV (full black line), $E_1=51.1$ MeV (dotted red line), $E_1=511$ MeV (dashed green line), $E_1=5.1$ GeV (dot-dashed blue line)  .\label{fig-ef-gamma}}
\end{figure}

The coordinate along the $x$ axis is chosen as $1-E_2/E_1$  and the results are presented in a log-log scale. For the lowest value considered for the incident electron energy,  the energy distribution decreases very fast with the ratio $\,E_2/E_1\,$: practically the entire distribution is contained in the interval $\,E_2\in(0.992E_1,E_1)\,$ which is an indication of the classicality of the process.   With the increase of the initial electron energy,  the energy spectrum is extended towards lower values of $\,E_2/E_1\,$, and its slope is much lower, as an indication of the onset of the quantum behaviour; for the largest value of the incident electron energy considered  ($\,\gamma_1=10^4$, $E_1=5.11\,$ GeV)  the electron  can lose up to 90\% of its energy.

The small peaks visible in the energy distribution for $\,\gamma_1=10^3$ and $\,\gamma_1=10^4$  correspond to the limit $\,E_a(1)\approx W_a(1)\,$ [Eq. (\ref{Wab})] of the energy range in which one photon absorption contributes.  For the other two values of $\,\gamma_1$ these points are located  at values of $\,1-E_2/E_1<10^{-4}\,$,  not represented in our figure.  The presence of this one photon peak can be explained  using the analogy  with the electron behaviour in  the monochromatic case, based on conservation laws valid in that case: for $\,E_2\,$ very close to $\,E_1\,$ the electron direction changes very little, and  the photon is emitted at an angle extremely small with respect  to the initial electron direction, i.e. $\,\theta_{k_2}\approx\pi$. It is  known  for a long time  \cite{Esarey} that for this geometry the terms with high $\,N\,$  in the radiation spectrum are suppressed, i.e. only the first few terms contribute to the total rate, even if $\,\eta\,$ is relatively large, and successive thresholds are visible in the electron distributions. For smaller values of $\,E_2\,$ the angular distribution widens, and many values of $\,N\,$ contribute to the total rate, i.e. the typical behaviour for large $\,\eta$ sets in: the successive maxima become broader and overlap, giving rise to a smooth continuum.

\section{Conclusions}\label{s-c}

The aim of our theoretical study of NLCS was a  first  description of several features of the scattered electron distributions  in the case of a pulsed electromagnetic wave.  
 We have identified two possible guides  for a qualitative understanding of the electron distribution: the monochromatic limit and the emitted radiation pattern. Which one is useful, if any, depends on several parameters: pulse shape (duration and intensity) and  initial electron momentum. While the role and condition of applicability of the first guide was identified  in the present study, the connection with the emitted radiation spectrum requires further investigation.

\appendix

\section{The solutions of equation (\ref{FN0})} \label{s-a}

In order to present  the properties of the solutions of the equation  (\ref{FN0}),   we use the notations (\ref{nota})
 and define   a set of dimensionless quantities:
\begin{equation}\label{uvX}
u_N=\frac{Q_N^0}{m_*c},\quad v_N=\frac{|{\bf Q}_N|}{m_*c},\quad X=\frac{|{\bf q}_2|}{m_*c}\ge0\,,
\end{equation}
 and we use the notations
\begin{equation}\label{alN}
\cos\alpha_N=Y\in[-1,1]\viq C_N=\frac{4u_N^2-(u_N^2-v_N^2+1)^2}{4v_N^2}\,.
\end{equation}
 It is useful to emphasize that while  $\,u_N, v_N\,$  and $\,C_N\,$ are determined by the initial conditions, $\,X\,$ and $\,Y\,$ are variables,  connected with the final momentum $\,\bfq_2\,$, which is subject to the condition (\ref{FN0}).

  Analyzing the equation (\ref{FN0}), we have  found that we have  to  distinguish between two cases:

 Case I: $\,u_N-v_N>1$, when $\,C_N<0$,

Case II:  $\,u_N-v_N<1$, when   $\,0<C_N<1\,$.

 We have found that if  the  momentum of the incident electron fulfills the condition $\,n_1\cdot q_1-m_*c>0$, we are in the case I for $\,N\hbar\,\omega_1>m_*c\,\frac{W_1-m_*c^2}{n_1\cdot q_1-m_*c}\,$ and in the case II for $\,N\hbar\,\omega_1<m_*c\,\frac{W_1-m_*c^2}{n_1\cdot q_1-m_*c}\,$. If $\,n_1\cdot q_1-m_*c<0$, we are in the case I for any value of $\,N$. 

The numerical examples presented in this paper (Sect. \ref{s-iiid} and \ref{s-iv}) refer to the case of head-on or nearly head-on collision of a very energetic electron with a laser pulse of moderate intensity ($\,\gamma_1\gg\eta\sim 1$). For these initial conditions we have $\,n_1\cdot q_1>m_*c$, and  we are in the case II for all values of $\,N$  which gives non-negligible contribution  to the electron distribution.

When written explicitly,  the function $\,F_N$  [Eq. (\ref{FN})] is a function of  the two unknowns $\,X$ and $\,Y$, defined in (\ref{uvX}) and (\ref{alN}), respectively. This way  Eq. (\ref{FN0}) becomes
\begin{equation}
 F_N(X,Y)\equiv\sqrt{X^2+1}-\sqrt{X^2+v_N^2-2\,X\,Yv_N}-u_N=0.
\end{equation}
It can be easily solved as an equation for $Y$, with the parameter $X$, leading to the expression
\begin{equation}
Y(X)=\frac{2\,u_N\,\sqrt{X^2+1}-(u_N^2+1-v_N^2)}{2\,X\,v_N}\,.\label{solY}
\end{equation}
 The properties of the function $\,Y(X)$  are different in the two cases mentioned before:

Case I  ($\;u_N>v_N+1\,$):  $\,Y\,$ increases  monotonously  with $\,X\,$ and the condition $\,|Y(X)|\le1\,$ leads to a domain of acceptable values of $\,X$
\begin{equation}
X\in\left[\,\vline\,\frac{u_N-v_N}2-\frac1{2(u_N-v_N)}\,\vline\,,\frac{u_N+v_N}2-\frac1{2(u_N+v_N)}\right]\label{domX}
\end{equation}

Case II  ($\;u_N<v_N+1\,$): $\,Y\,$ has a minimum
\begin{equation}\label{Ymin}
Y_{\mathrm{min}}=\sqrt{C_N}\in(0,1)\,,
\end{equation}
reached for 
\begin{equation}\label{xd}
X_{\mathrm{d}}=\frac{\sqrt{4u_N^2-(u_N^2+1-v_N^2)^2}}{u_N^2-v_N^2+1}\,.
\end{equation}
 The condition $\,|Y(X)|\le1\,$ leads to the same domain (\ref{domX}) of values for $\,X\,$ as in the case I.
When expressed in terms of energy of the dressed electrons, the interval (\ref{domX}) is $\,W_2\in[W_a,W_b]\,$ with 
\begin{equation}
W_a(N)={\cal E}_N-\frac{N\hbar\omega_1}{{\cal E}_N/c-\mid {\bf Q}_N\mid}\,n_1\cdot q_1,\quad W_b(N)={\cal E}_N-\frac{N\hbar\omega_1}{{\cal E}_N/c+\mid {\bf Q}_N\mid}\,n_1\cdot q_1\,.\label{Wab}
\end{equation}

\vspace{0.15cm}

Going the other way around, i.e. solving the equation (\ref{FN0}) for the unknown $\,X$ as function of $\,Y$,   we  find:

 i) only one solution in the case I, namely  
\begin{equation}
X_+(Y)=\frac{v_NY(u_N^2-v_N^2+1)+2u_N\,v_N\,\sqrt{Y^2-C_N}}{2(u_N^2-v_N^2Y^2)}\label{solxi}
\end{equation}
for any $Y\in[-1,1]$,

 ii)   two values for $\,X$ in case II,
\begin{equation}
X_{\pm}(Y)=\frac{v_N\,Y\,(u_N^2-v_N^2+1)\pm2u_Nv_N\sqrt{Y^2-C_N}}{2(u_N^2-v_N^2\,Y^2)}\label{solxii}
\end{equation}
for $\,Y\in[Y_{\mathrm{min}},1]$.  At $\,Y=Y_{\mathrm{min}}\,$ the two solutions $\,X_\pm\,$ coalesce to the value $\,X_{\mathrm{d}}$  in Eq. (\ref{xd}).

 In conclusion we have established  the equations that describe the position of the singularities brought by the $\,\delta$-functions in (\ref{distr}) in terms of the variables $\,X\,$ or $\,Y\,$.

\section{Study of the solutions of Eq. (\ref{eqtp})}\label{s-b}
With the notations 
\begin{equation}
u_{q_2}=\cos\theta_{q_2},\quad y=\cos\theta_N,\quad \delta\phi=\phi_{q_2}-\phi_N\,,
\end{equation}
Eq. (\ref{eqtp}) for $\,\cos\delta\phi\,$ becomes
\begin{equation}
Y=u_{q_2}y-\sqrt{1-u_{q_2}^2}\sqrt{1-y^2}\cos\delta\phi\,.
\end{equation}
 The parameters $\,u_{q_2},y,Y\,$ are subject to the conditions $|Y|,|y|,|u_{q_2}|\le1$.
The  solution
\begin{equation}
\delta\phi=\pm\arccos\rho(u_{q_2}),\quad\rho(u_{q_2})=\frac{Y-u_{q_2}y}{\sqrt{1-u_{q_2}^2}\sqrt{1-y^2}}.
\end{equation}
is acceptable, if $\rho$ has the modulus less than unit. From the expression of  its  derivative 
\begin{equation}
\frac{d\rho}{du_{q_2}}=\frac{Yu_{q_2}-y}{(1-u_{q_2}^2)^{3/2}\sqrt{1-y^2}}
\end{equation}
we see that  for $\,|Y|<|y|$,  $\,\rho\,$ is a monotonic function, taking values  between $-1$ and $1$ when $\,u_{q_2}\,$ takes values in the interval 
\begin{equation}
u_{q_2}\in[\,u_{q_2,{\mathrm{min}}}=yY-\sqrt{1-y^2}\sqrt{1-Y^2},
\;u_{q_2,\mathrm{max}}=yY+\sqrt{1-y^2}\sqrt{1-Y^2}\,]\,.\label{intx}
\end{equation}
In terms of angles this condition   becomes Eq. (\ref{condte}). If  $\,|Y|>|y|\,$, then $\,\rho(u_{q_2})\,$ has an extremum equal to   $\,{\mathrm{sgn}}(Y)\sqrt{\frac{Y^2-y^2}{1-y^2}}$ for $\,x=Y/y\,$; the condition  $\,|\rho(u_{q_2})|<1\,$ leads to the same interval (\ref{intx}).

 When  equation (\ref{eqtp})  is solved for the unknown $\,u_{q_2}\,$  at fixed  $\,\phi_{q_2}$, we obtain: For $\,|Y|>|y|,$ there are two solutions 
\begin{equation}\label{solecuua}
u_{q_2}^{(\pm)}=\frac{Yy\pm\sqrt{1-y^2}|\cos\delta\phi|\sqrt{y^2+(1-y^2)\rho^2-Y^2}}
y^2+(1-y^2)\cos^2\delta\phi
\end{equation}
acceptable only for $\,\cos\delta\phi\in[\,\sqrt{\frac{Y^2-y^2}{1-y^2}},1\,]\,$ if $\,Y>0$, and for $\,\cos\delta\phi\in[-1,-\sqrt{\frac{Y^2-y^2}{1-y^2}}]\,$ if $\,Y<0$.  For  $\,|Y|<|y|\,$ there is only one solution for any $\,\phi\in[-\pi,\pi]$: $\,u_{q_2}^{(-)},$ if $\,y\,\cos\delta\phi>0$, or $\,u_{q_2}^{(+)}\,$ if $\,y\,\cos\delta\phi<0$.

Now we can write explicitly  the domain $\,{\cal D}(C_N)\,$ introduced in Section \ref{s-iiia}, defined by the condition $\,\cos\alpha_N\le \sqrt{C_N}$. We  refer to  the solutions (\ref{solecuua}), assuming that in their expression $\,Y\,$ was replaced by $\,\sqrt{C_N}>0$. Then the domain ${\cal D}(C_N)$ can be   described by:

 $\,\phi_{q_2}\in[\phi_N-\phi_0,\phi_N+\phi_0]$, $\,\cos\theta_{q_2}\in[\,u_{q_2}^{(-)},u_{q_2}^{(+)}\,]$, where $\,\phi_0=\arccos\sqrt{\frac{C_N-\cos^2\theta_N}{\sin^2\theta_N}}\,$, for $\,|\cos\theta_N|\le\sqrt{C_N}$;

$\phi_{q_2}\in[\phi_N-\pi,\phi_N+\pi]$, $\cos\theta_{q_2}\in[u_{q_2}^{(-)},1]$ if $\cos(\phi_{q_2}-\phi_N)>0$ and $\cos\theta_{q_2}\in[u_{q_2}^{(+)},1]$ if $\cos(\phi_{q_2}-\phi_N)>0$, for $\cos\theta_N>\sqrt{C_N}$;

$\phi_{q_2}\in[\phi_N-\pi,\phi_N+\pi]$, $\cos\theta_{q_2}\in[-1,u_{q_2}^{(-)}]$ if $\cos(\phi_{q_2}-\phi_N)>0$ and $\cos\theta_{q_2}\in[-1,u_{q_2}^{(+)}]$ if $\cos(\phi_{q_2}-\phi_N)>0$ for $\cos\theta_N<-\sqrt{C_N}$.

\acknowledgments

This work was supported by CNCSIS-UEFISCSU, project number 488 PNII-IDEI 1909/2008.  M.B. acknowledges the support of the strategic grant POSDRU/89/1.5/S/58852, Project ``Postdoctoral programme for training scientific researchers'' cofinanced by the European Social Found within the Sectorial Operational Program Human Resources Development 2007-2013. V.D. thanks  A. Ilderton for  a useful discussion about the description of the initial state of the electron.


\begin{thebibliography}{14}
\bibitem{NNR} N. B. Naroznhyi, A.I. Nikishov, and V. I. Ritus, JETP {\bf 47}, 930 (1964) [Sov. Phys. JETP {\bf 20}, 622 (1965)].
\bibitem{SaHa} Y. I. Salamin, S. X. Hu, K. Z. Hatsagortsyan and C. H. Keitel, Physics Reports {\bf 427}, 41 (2006).
\bibitem{KrKa} F. Ehlotzky, K. Krajewska and J. Z. Kaminski,  Rep. Prog. Phys. {\bf 72}, 046401 (2009).
\bibitem{E144-1} C. Bula {\it et al.}, Phys. Rev. Lett {\bf 76}, 3116 (1996). 
\bibitem{E144-2} C. Bamber {\it et al.}, Phys. Rev. D {\bf 60}, 092004 (1999).
\bibitem{RMP} A. Di Piazza, C. M\"uller, K. Z. Hatsagortzyan and C. H. Keitel, arXiv:1111.3886v2 [hep-ph], accepted by Reviews of Modern Physics.
\bibitem{RR}  A. Di Piazza, K. Z. Hatsagortsyan, and C. H. Keitel,  Phys. Rev. Lett. {\bf 105}, 220403 (2010).
\bibitem{Sok} I. V. Sokolov {\it et al.}, Phys. Rev. E {\bf 81}, 036412 (2010).
\bibitem{BF} M. Boca and V. Florescu, Phys. Rev. A {\bf 80}, 053403 (2009).
\bibitem{altii1} T. Heinzl, D. Seipt, and B. K\"ampfer, Phys. Rev. A {\bf 81}, 022125 (2010).
\bibitem{altii2} D. Seipt and B. K\"ampfer, Phys. Rev. A {\bf 83}, 022101 (2011).
\bibitem{altii3} F. Mackenroth and A. Di Piazza, Phys. Rev. A {\bf 83}, 032106 (2011).
\bibitem{altii4} F. Mackenroth, A. Di Piazza, and C. H. Keitel, Phys. Rev. Lett. {\bf 105}, 063903 (2010).
\bibitem{AI} A. Ilderton, Phys. Rev. Lett. {\bf 106}, 020404 (2011).
\bibitem{altele} T. Heinzl, A. Ilderton, M. Marklund, Phys. Lett. B {\bf  692},  250 (2010).
\bibitem{size} J. Gao, J. Phys. B: At. Mol. Opt. Phys. {\bf 39}, 1345 (2006).
\bibitem{galkin} A. L. Galkin {\it et al}, Contrib. Plasma Phys. {\bf 49}, 593 (2009).
\bibitem{Th-exp} M. Babzien {\it et al.} Phys. Rev. Lett. {\bf 96}, 054802 (2006).
\bibitem{int}  The White Book of ELI Nuclear Physics, Bucharest-Magurele, Romania, http://www.eli-np.ro/documents/ELI-NP-WhiteBook.pdf.
\bibitem{arki}A. Di Piazza, C. M\"uller, K. Z. Hatsagortsyan, and C. H. Keitel, 
 arXiv:1111.3886v1 [hep-ph].
\bibitem{CP1} J. P. Corson, J. Peatross, C. Muller, K. Z. Hatsagortsyan, Phys. Rev. A {\bf 84}, 053831 (2011).
\bibitem{CP2} J. P. Corson and J. Peatross, Phys. Rev. A {\bf 84}, 053832 (2011).
\bibitem{BDF} M. Boca, V. Dinu and V. Florescu, Nucl. Instrum. Meth. B, {\bf 279},  12 (2012).
\bibitem{qdist}  Y. I. Salamin and F. H. M. Faisal, Phys. Rev. A {\bf 54}, 4383 (1996).
\bibitem{PhysSc} M. Boca and A. Oprea, Phys. Scr. {\bf 83}, 055404 (2011).
\bibitem{Esarey} E. Esarey, S. K. Ride, P. Sprangle, Phys. Rev. E {\bf 48}, 3003 (1993).

\end{thebibliography}
\end{document}